\documentclass[aps,twocolumn,preprintnumbers,showpacs,showkeys]{revtex4}
\usepackage{epsfig}
\usepackage{amssymb,amsmath,amsfonts,amsthm,graphicx}
\graphicspath{{./Figures/}}                 
\def\Pol{{\mathcal{P}}}
\def\diag{{\rm diag}}

\newcommand{\beq}{\begin{equation}}
\newcommand{\eeq}{\end{equation}}
\newcommand{\bea}{\begin{array}}
\newcommand{\eea}{\end{array}}
\newcommand{\beqa}{\begin{eqnarray}}
\newcommand{\eeqa}{\end{eqnarray}}
\newcommand{\Fig}[1]{Fig.~\ref{#1}}

\newcommand{\tr}{\operatorname{Tr}}

\newcommand{\half}{{\scriptstyle{\frac{1}{2}}}}

\def\cA{{\cal{A}}}
\def\cT{{\hat{\beta}}}
\def\cT{{\beta}}

\def\beqa{\begin{eqnarray}}
\def\eeqa{\end{eqnarray}}
\def\pl{{{\cal P}_\infty}}
\def\plo{{{\cal P}_\infty^0}}
\def\tr{{\rm tr}}
\def\Tr{{\rm Tr}}

\begin{document}

\preprint{ITEP-LAT/2013-11,~HU-EP/13-51 (revised)}

\title{Topology near the transition temperature in lattice gluodynamics \\ 
analyzed by low lying modes of the overlap Dirac operator}

\author{E.-M. Ilgenfritz}
\affiliation{Joint Institute for Nuclear Research, VBLHEP, 141980 Dubna, Russia}

\author{B.~V.~Martemyanov}
\affiliation{
Institute of Theoretical and Experimental Physics, 117259 Moscow, Russia\\
National Research Nuclear University MEPhI, 115409, Moscow, Russia\\
Moscow Institute of Physics and Technology, 141700, Dolgoprudny, Moscow Region,
Russia}

\author{M.~M\"uller-Preussker}
\affiliation{Humboldt-Universit\"at zu Berlin, Institut f\"ur Physik,
  12489 Berlin, Germany}

\date{February 3, 2014}

\begin{abstract}
Topological objects of $SU(3)$ gluodynamics are studied at the infrared scale 
near the transition temperature with the help of zero and near-zero modes of 
the overlap Dirac operator.
We construct UV filtered topological charge densities corresponding to three 
versions of the temporal boundary condition applied to this operator, for which 
the zero mode is known to be located on corresponding three constituent dyons 
(antidyons) in the reference case of an analytical (anti)caloron solution. 
The clustering of the three topological charge densities marks the positions 
of three types of dyons and antidyons which can therefore be considered as 
present in equilibrium (Monte Carlo) gluonic fields at the given resolution 
scale. We classify them either as constituents of nondissociated (anti)calorons 
or as constituents of (anti)dyon pairs or as isolated (anti)dyons. 
The pattern of the Polyakov loop describing the centers and the interior of
these clusters is observed after a limited number of overimproved cooling 
steps and resembles the description known from analytical caloron solutions.
\end{abstract}



\maketitle

\section{Introduction}
\label{sec:introduction}

Two basic properties of QCD are confinement (of quarks and gluons) and the
spontaneous breaking of chiral symmetry at low temperature and density.
Both properties are believed to be intimately connected with each other and
to originate from a certain complex structure of the QCD vacuum state, the 
simplest manifestations of which being condensates of gluon and quark fields. 
The field fluctuations contributing to these condensates are, however,
space-time and scale dependent. One of the aims of Lattice Gauge Theory 
is to reveal the corresponding structures. One school of thought claims 
that - at the infrared scale - the origin of both mechanisms can be 
traced back to semiclassical objects of QCD. 
Until the late 1990's lattice studies tried to figure out the ``instanton 
structure'' of the vacuum by ``cooling'' or ``smearing'', the role of which
today is taken by the Wilson flow~\cite{ Luscher:2010iy,Luscher:2013vga}. 
The number of cooling steps was providing the running scale of 
resolution~\cite{GarciaPerez:1998ru}. For reviews of the instanton picture 
itself we refer to the reports~\cite{Schafer:1996wv,Diakonov:2002fq}.
Then, including a slight extension to non-zero temperature $T$, more general 
semiclassical objects, Kraan-van Baal-Lee-Lu (KvBLL) calorons~\cite{Kraan:1998pm,
Kraan:1998sn,Lee:1998bb}, have replaced instantons as the carriers of 
topological charge, with new features like a variable asymptotic holonomy 
and (topologically) fractionally charged constituents (dyons). Still, 
KvBLL calorons have not reached that level of familiarity that instantons 
have enjoyed among lattice practitioners. Today it is folklore to say 
that the instanton gas is able to explain chiral symmetry breaking while
it fails to provide a mechanism for confinement. Constituent dyons, 
however, when considered as rarefied gas without interaction or with 
Coulomb-like interaction, give confining behavior for space-like Wilson 
loops and for correlators of Polyakov loops. The history of this idea 
ranges from the 70's to the recent past~\cite{Polyakov:1976fu,
Martemyanov:1997ks,Gerhold:2006sk,Diakonov:2007nv,Bruckmann:2011yd}.

Therefore, it is of some interest to search for dyons in standard Monte 
Carlo configurations, representing lattice gauge fields at different 
temperatures, in order to assess the reality of these models. For the 
purpose of this investigation, the fermion eigenmodes of a chirally 
(almost) perfect Dirac operator (i. e. the massless overlap operator) 
are promising to be an adequate tool. In this paper we shall employ 
this tool to pure $SU(3)$ gauge theory. We stress that a cutoff with 
respect to the eigenvalues plays the role of the running scale. This 
interpretation has been given to the cutoff already in previous 
applications of the overlap operator~\cite{Ilgenfritz:2007xu} for 
investigations of topological structure at $T=0$.
In the context of finite temperature the splitting of calorons into 
dyons is a phenomenon worth to study. Then the choice of the temporal 
boundary condition (b.c.) applied to the analysing Dirac operator is an
additional free parameter allowing to separate different degrees 
of freedom inside a caloron (dyon constituents) according to their 
inherent holonomy.

The caloron with nontrivial holonomy~\cite{Kraan:1998pm,Kraan:1998sn,Lee:1998bb} 
has the remarkable property that the single zero mode of the Dirac operator 
locates on different constituent 
dyons~\cite{GarciaPerez:1999ux,Chernodub:1999wg}, 
depending on the temporal b.c. applied to the Dirac operator.
In lattice simulations, the change of zero mode localization with the change 
of b.c. was observed for thermal configurations of lattice gauge 
fields~\cite{Gattringer:2002wh,Gattringer:2002tg}. In the case of $SU(2)$ 
lattice gauge theory it has been seen that this property is shared also by 
near-zero modes~\cite{Bornyakov:2007fm,Bornyakov:2008im}. Thus, the low lying 
modes of the overlap Dirac operator can be used as an effective tool to 
detect distinct topological objects at a corresponding scale directly within 
Monte Carlo configurations of lattice gauge fields (without cooling or 
smearing). 

The angle characterizing the b.c. needs to be varied in order to detect 
really {\it all sorts of dyons} by the fermionic method. Only this allows 
to check the phenomenological relevance of the KvBLL caloron picture which 
is assumed in models like those mentioned above. In $SU(3)$ lattice gauge 
theory, the properties of constituent dyons have been studied thoroughly 
earlier in cooled (i. e. classical) configurations of lattice gauge 
fields~\cite{Ilgenfritz:2005um}. 
The interrelation between localization and delocalization (in space-time 
position and with respect to extension) and the angle of the b.c. of the 
fermion field was demonstrated in detail. The phenomenological relevance of 
the dyon picture, however, was not yet in the scope of that work.

To carry over the method of analysis, developed previously for the $SU(2)$ case
in Refs.~\cite{Ilgenfritz:2006ju,Bornyakov:2007fm,Bornyakov:2008im}, to the 
case of $SU(3)$ gauge fields is not trivial. Having three (and among them, 
two independent) eigenphases of each local holonomy, enforces to consider 
the complex plane of the Polyakov loop and to introduce an angle 
modifying the b.c. beyond the antiperiodic and periodic case.
Our aim is to classify, according to the KvBLL caloron picture (outlined in the 
Appendix), topological objects in field ensembles of pure $SU(3)$ gluodynamics,
generated close to the deconfinement transition temperature, with the help
of zero and near-zero modes of the overlap Dirac operator, modified by 
different temporal boundary conditions.

Our previous study of topological objects at non-zero temperature in $SU(2)$ 
lattice fields, based on smearing~\cite{Ilgenfritz:2004zz,Ilgenfritz:2006ju} 
or on an overlap fermion analysis~\cite{Bornyakov:2007fm,Bornyakov:2008im},
has resulted in the following picture of the topological content of $SU(2)$ 
gauge theory. At low temperatures, the relevant topological objects are 
nondissociated calorons with maximally nontrivial holonomy. With increasing 
temperature, their composite nature becomes recognizable as the dissociation 
into two dyons (i. e. selfdual non-Abelian) monopoles of topological charge 
$\pm 1/2$. Approaching the critical temperature $T_c$ (of the deconfining 
phase transition), approximately half of the calorons have become dissociated, 
retaining the symmetry of the two dyons. Above the critical temperature a 
non-zero expectation value of the averaged Polyakov loop $\langle L \rangle$ 
develops. This results in an asymmetry between the dyons: light dyons (with 
the local Polyakov loop of same sign as $\langle L \rangle$) become the most 
abundant topological objects, while heavy dyons (and, even more so, 
nondissociated calorons) are suppressed.
In the present paper we will see to what extent this picture translates
to the three dyons per caloron of $SU(3)$ gauge theory, which would be 
distingushed -- by analogy -- by the phase of the complex-valued Polyakov loop. 

In Section~\ref{sec:thermal-ensembles} the generation of the two lattice 
ensembles characterizing temperatures slightly above and below the
deconfining transition is described. 
In Section~\ref{sec:clusters} the dependence of the eigenvalue spectrum of the 
overlap operator on the observed physical phase {\it and on the boundary condition} is 
demonstrated. We sketch the fermionic construction of the topological charge 
density, restricted to the infrared scale by the cutoff, with the help of the 
eigenmodes for three different fermionic boundary conditions. Finally, the 
results of a cluster analysis of these densities is presented with respect 
to the formation of clusters (``dyons'') of different local holonomy and of 
``bound states'' of different type dyons (including complete calorons).
Section~\ref{sec:polyakov-portrait} raises the question to what extent the 
fermionic topological density (if averaged over the three different b.c.'s 
in order to include all types of dyons) can be reproduced by cooling 
techniques. An optimal number of cooling steps is found which solves this 
problem in an impressive way. With this amount of cooling, it becomes 
possible to understand the systematics of the Polyakov loop in different
single dyons and to understand the scatter plot of the Polyakov loop inside 
bigger clusters (representing dyon pairs and calorons). 
In Section~\ref{sec:conclusions} we conclude that the resolution chosen for 
this investigation supports the idea that semiclassical objects of dyon type
characterize the topological structure at this scale.
The Appendix is devoted to a brief summary of those aspects of calorons that 
have been essential for this study.

\section{Setup of the investigation}
\label{sec:thermal-ensembles}

The $SU(3)$ gauge field configurations for this investigation
have been generated on a lattice of size $20^3 \times 6$  
by sampling the pure $SU(3)$ gauge theory using the 
L\"uscher-Weisz action ~\cite{Luscher:1984xn}. 
Improved gauge actions are known to be mandatory for analyses using the overlap 
Dirac operator, in order to take full advantage of their good chiral properties.
The sampled gauge fields are smoother than those sampled with the Wilson action.
In particular, the idea of our analysis rests on the observation that changing 
the boundary condition leaves the number of zero modes unchanged. 
The L\"uscher-Weisz action has also been used in the QCDSF topological 
studies~\cite{Ilgenfritz:2007xu} of pure Yang-Mills theory with overlap fermions
and by Gattringer and coworkers~\cite{Gattringer:2001jf,Gattringer:2001ia} 
when they were using a specific chirally improved fermion action for topological 
investigations. In the $SU(2)$ case, the tadpole-improved Symanzik action 
has been applied for analogous reasons in our previous 
work~\cite{Bornyakov:2007fm,Bornyakov:2008bg,Bornyakov:2008im}.

In addition to the plaquette term (pl), the L\"uscher-Weisz action includes 
a sum over all $2 \! \times \! 1$  rectangles (rt) and a sum over all 
parallelograms (pg), i.e.~all possible closed loops of length 6 along the 
edges of all 3-cubes
\beqa
S[U]  = & \beta & \left(\sum_{pl} \frac{1}{3}  \mbox{Re~Tr}  [ 1 - U_{pl} ]\right.
\nonumber 
\\ 
& + &  c_1 \sum_{rt} \frac{1}{3}  \mbox{Re~Tr}  [ 1 - U_{rt} ] 
\\
& + & 
\left. c_2 \sum_{pg} \frac{1}{3}  \mbox{Re~Tr}  [ 1 - U_{pg} ] \right)\,,
\nonumber 
\label{sgauge}
\eeqa
where $\beta$ is the principal inverse coupling parameter, while the coefficients
$c_1$ and $c_2$ are computed using results of one-loop perturbation theory and 
tadpole improvement~\cite{Luscher:1985zq,Snippe:1997ru,Lepage:1992xa}:
\beq
c_1 =  -  \frac{1}{ 20  u_0^2}  
[ 1 + 0.4805  \alpha ]\,,~~
c_2 =  -  \frac{1}{u_0^2}  0.03325  \alpha \,.
\eeq
For a given $\beta$, the tadpole factor $u_0$  and the lattice coupling constant
$\alpha$ are self-consistently determined in terms of the average plaquette 
\beq
u_0  =  \Big( \langle \frac{1}{3} \mbox{Re~Tr}~U_{pl} \rangle 
\Big)^{1/4}\,, \quad \alpha  =  -
\frac{ \ln \Big( \langle \frac{1}{3} \mbox{Re~Tr}~U_{pl} \rangle 
\Big)}{3.06839} 
\eeq 
in the course of a  series of iterations.

Two ensembles, each consisting of 50 configurations, have been generated for 
the overlap analysis with values of the principal inverse coupling values 
$\beta = 8.20$ and $\beta = 8.25 $. For the generation of these configurations 
1500 empty sweeps have been performed in between in order to make the 
configurations uncorrelated among each other. According to previous 
work~\cite{Gattringer:2002mr}, these ensembles correspond to temperatures 
close to the phase transition point $\beta=\beta_c$. 

By lowering $\beta$ we have tried to look deeper into the confining phase in 
order to find more signatures for caloron dominance (with equally weighted 
constituent dyons related to maximally non-trivial holonomy). However, as 
mentioned above, the proper applicability of the overlap operator 
diagonalization is limited to a range of sufficiently large $\beta$-values 
(i. e. to sufficiently smooth lattice gauge fields) even if the 
L\"uscher-Weisz action is in use. 
This closeness to $\beta_c$ had to be accepted as necessary condition in the 
confining phase in order to ensure that the number of zero-modes 
(the ``topological index'') of a given gauge field will not change with the 
choice of boundary condition for the fermion field.      

\begin{figure*}[!htb]
\centering
\vspace{1cm}
\centering
a)\hspace{0.3cm} \includegraphics[width=.37\textwidth]{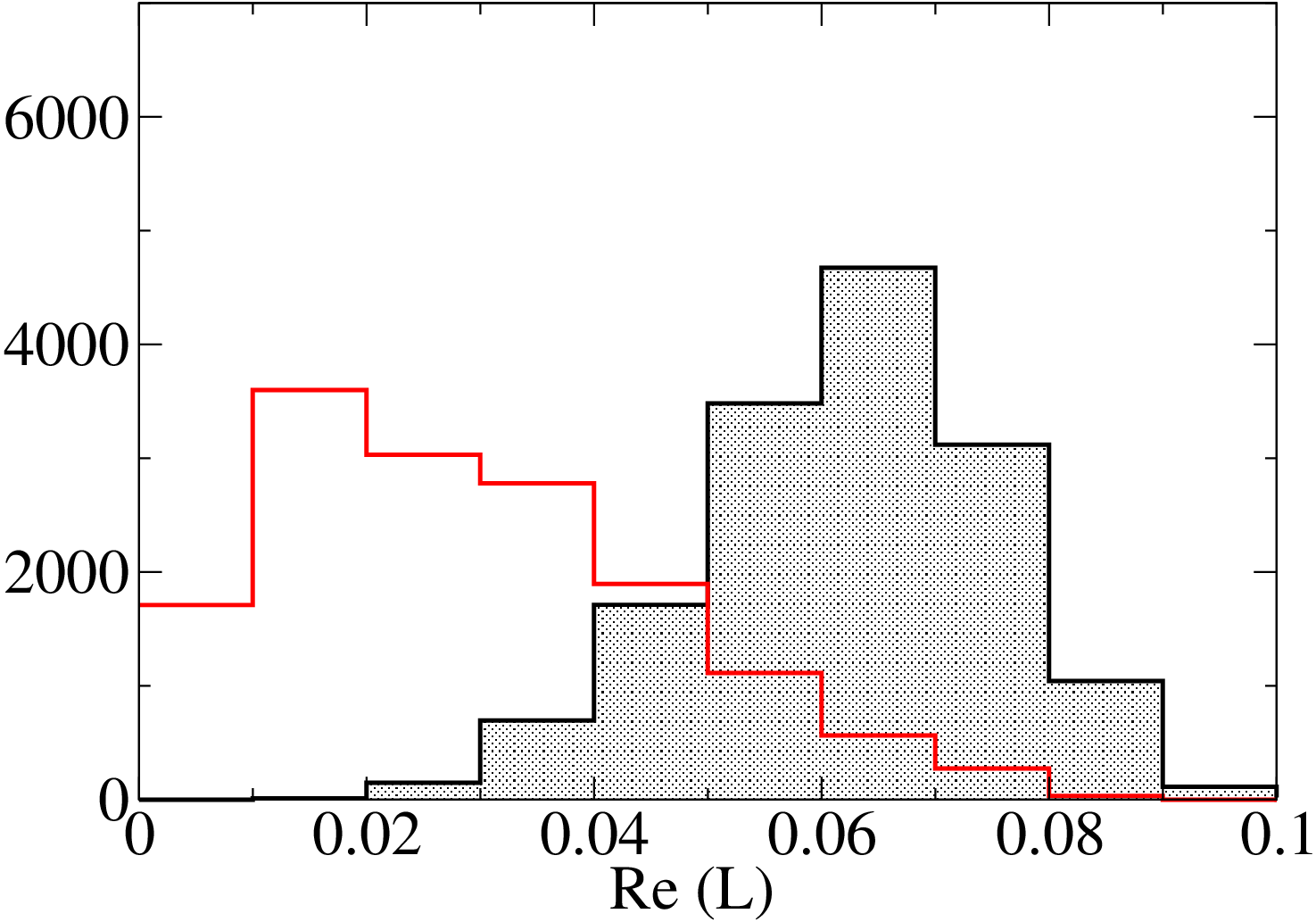}%
\hspace{0.3cm}
b)\hspace{0.3cm} \includegraphics[width=.37\textwidth]{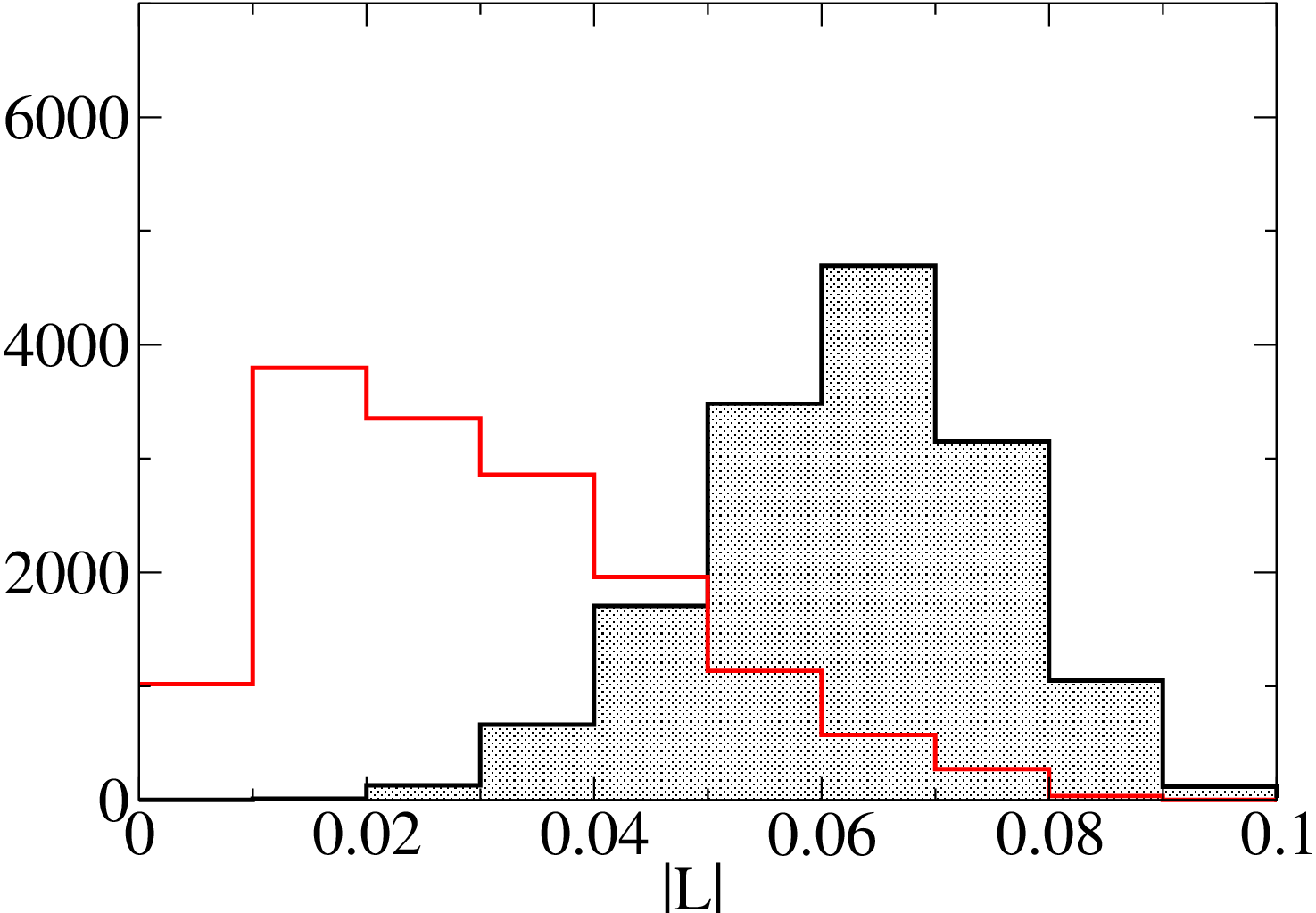}\\
\caption{The distributions a) of the real part of the spatially averaged 
Polyakov loop $Re(L)$ (eventually rotated to the real $Z_3$ sector)
and b) of the modulus $|L|$, both at $\beta=8.20$ (red curve) and 
$\beta=8.25$ (shadowed), each based on a statistics of 15000 
configurations.}
\label{fig:pldistr}
\end{figure*}

In order to characterize the two ensembles, we present in 
Fig. \ref{fig:pldistr}a the distributions of the real part of the spatially 
averaged Polyakov loop $Re(L)$ (without any smearing or cooling) 
rotated to the real $Z_3$ sector $-\pi/3 < \arg(L) < \pi/3$. A global $Z_3$ 
rotation has been applied whenever $L$ was found with 
$  \pi/3 < \arg(L) <   \pi$ or  
$- \pi/3 > \arg(L) > - \pi$.
The distributions have been obtained from 15000 subsequent Monte Carlo 
configurations. They were sampled independently from the configurations 
that have been used in the overlap analysis lateron. 
We show the analogous comparison for the distributions of the modulus of 
the spatially averaged Polyakov loop $|L|$ in Fig. \ref{fig:pldistr}b.
Both kinds of distributions suggest that configurations obtained 
at $\beta=8.20$ belong mainly to the confining phase (with an 
admixture from the deconfining phase), and that the configurations 
at $\beta=8.25$ refer exclusively to the deconfining phase.

\section{Topological clusters}
\label{sec:clusters}

We have analyzed the configurations of the two ensembles 
by identifying and investigating in each case $N = 20$ near-zero eigenmodes 
of the overlap Dirac operator. Its realization is an $SU(3)$ extension 
of the $SU(2)$ overlap Dirac operator described in~\cite{Bornyakov:2007fm}. 
Originally, the choice of $N=20$ modes was mainly dictated by the limited 
amount of workstation resources for this long-term project. As described later, 
the resulting analyzing power happened to be tantamount to approximately
20 overimproved cooling steps. The spectral analysis has been performed 
for three temporal b.c.'s applied to the fermion field $\psi$, for which 
-- in the case of a single-caloron solution with maximally nontrivial 
holonomy -- the zero mode of the fermion field is maximally localized at 
one of its three constituent dyons (see Appendix):
\beq 
\psi (1/T) = \exp(i\phi)\psi(0)
\label{eq:bc1}
\eeq 
with
\beq
\phi = \left\{
\begin{array}{ll}
 \phi_1  \equiv -\pi/3\,, \\
 \phi_2  \equiv +\pi/3\,, \\
 \phi_3  \equiv ~~~~\pi\,. \\
\end{array} 
\right.
\label{eq:bc2}
\eeq
For each of these b.c.'s we have determined the topological index and have 
checked that it was obtained -- in the $\beta$ range under consideration -- 
independent of the choice of $\phi$. 

\begin{figure*}[!htb]
\centering
a)\hspace{0.3cm} \includegraphics[width=.37\textwidth]{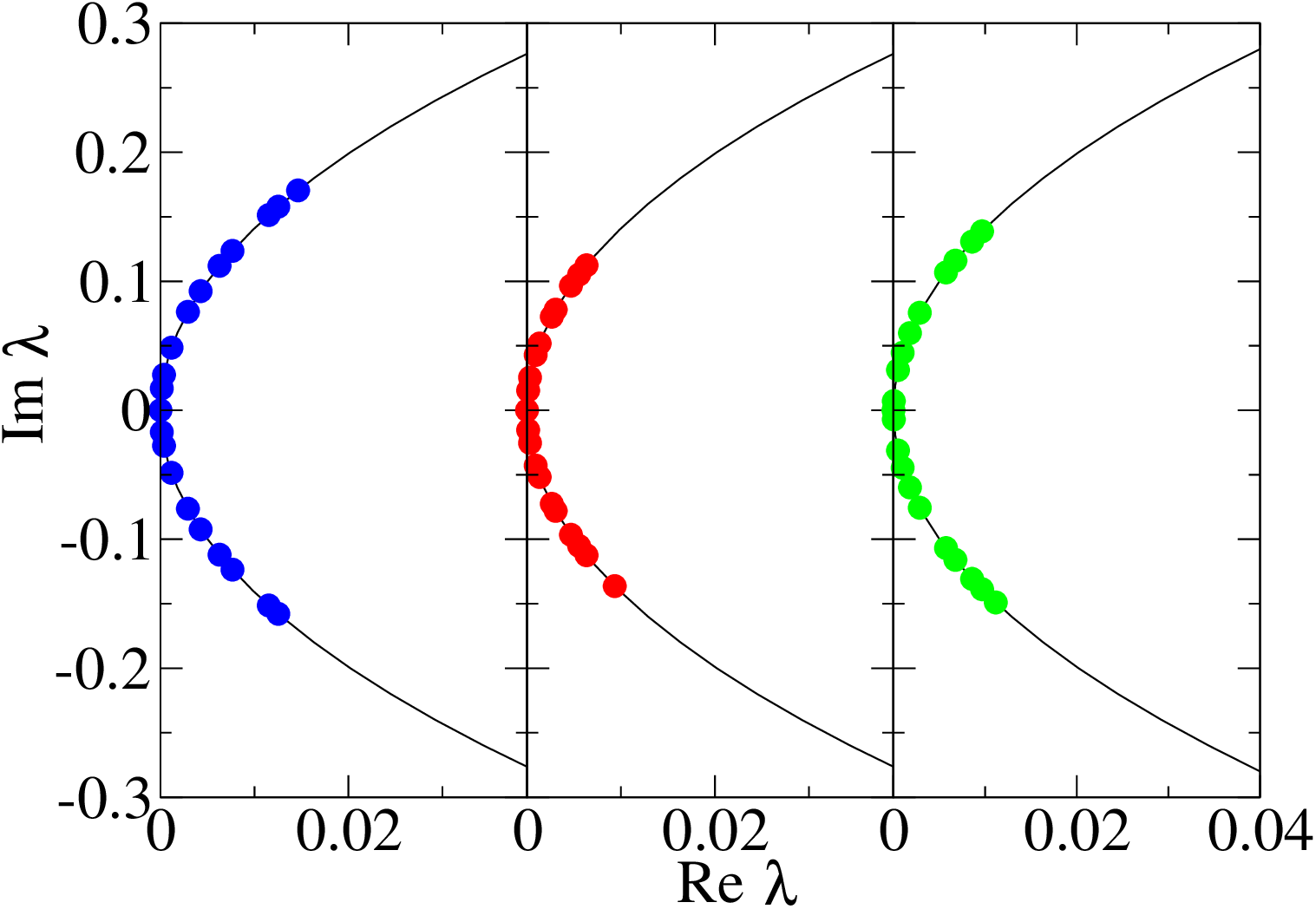}%
\hspace{0.3cm}
b)\hspace{0.3cm} \includegraphics[width=.37\textwidth]{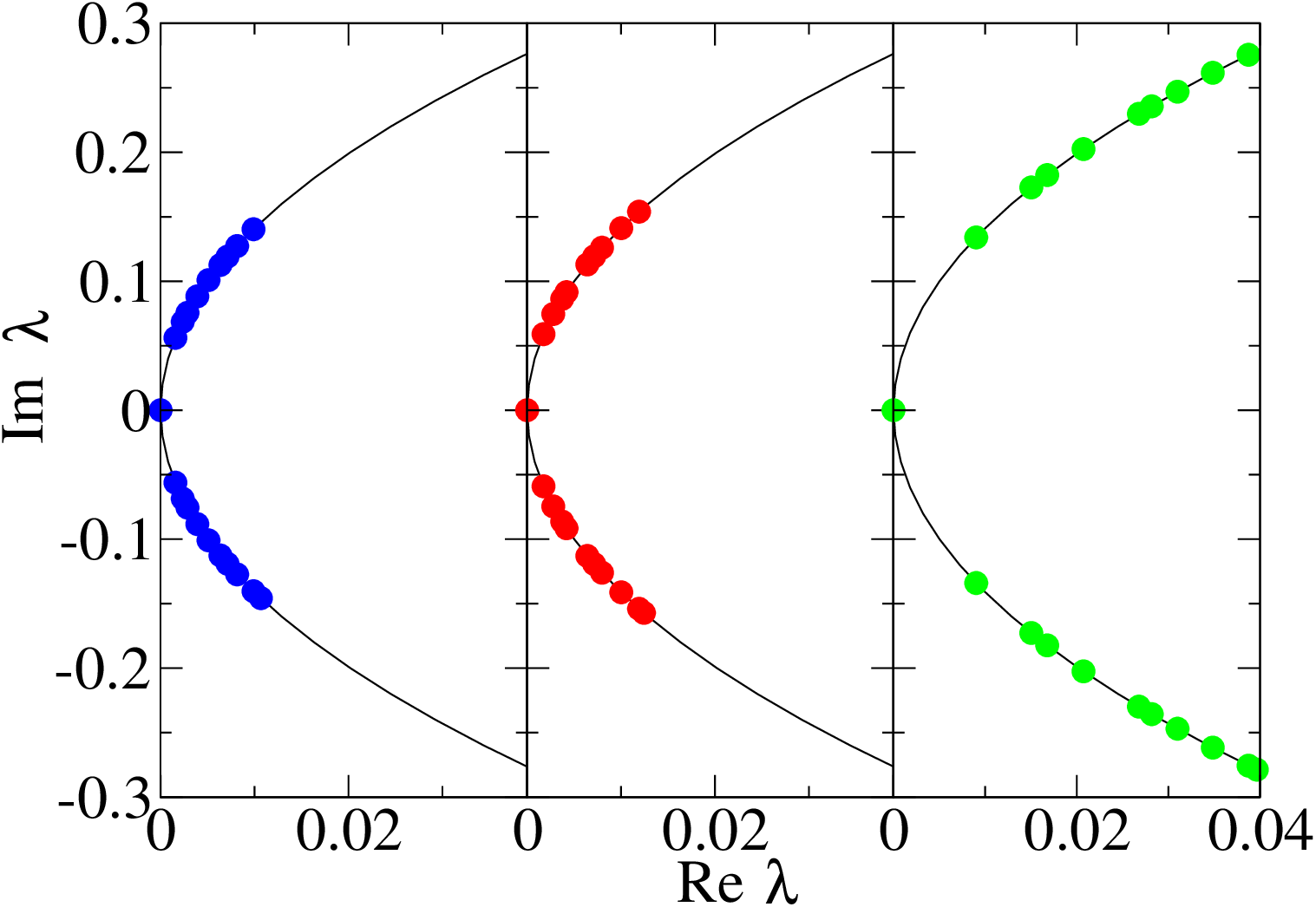}\\
\caption{For two typical configurations and in each case for the three 
versions of the fermion temporal b.c. (according to Eqs. 
(\ref{eq:bc1}, \ref{eq:bc2})) the 20 lowest eigenvalues of the $SU(3)$ 
overlap Dirac operator are shown, for a) $\beta=8.20$  and b) $\beta=8.25$, 
respectively. In both of the panels for $\phi=\phi_1$ the spectrum is 
shown left (in blue), for $\phi=\phi_2$ in the middle (in red), and 
for $\phi=\phi_3$ (antiperiodic b.c.) it is plotted right (in green), 
respectively.}
\vspace{1cm}
\label{fig:spectra}
\end{figure*}

The corresponding spectra of the 20 near-zero eigenmodes are shown 
in Fig. \ref{fig:spectra} for two configurations representative for 
$\beta=8.20$ and $\beta=8.25$, respectively. 
In the confinement case (a) these spectra have a nonzero density around 
zero value (signalling spontaneous violation of chiral symmetry); in the  
deconfinement case (b) a gap in the spectra is appearing. 
Notice for the latter case the occurence of an even larger gap width 
for the third (i.e. antiperiodic) boundary condition ($\phi_3$). The spatially 
averaged Polyakov loop $L$ for this configuration has fallen into the real 
$Z_3$ sector $-\pi/3 < \arg(L) < \pi/3$. The observation of the gap is
in agreement with results of earlier studies ~\cite{Gattringer:2002dv} 
using the chirally improved Dirac operator.
In order to proceed further, we have reconstructed from the zero and the
non-zero modes the profiles of the UV-filtered topological charge density 
according to its spectral representation (for details 
see \cite{Hasenfratz:1998ri,Ilgenfritz:2007xu})
\beq
q_{i,N}(x) = - \sum_{j=1}^N 
\left( 1 - \frac{\lambda_{i,j}}{2} \right) \psi^{\dagger}_{i,j}(x)\gamma_5 
\psi_{i,j}(x)\,,
\label{eq:truncated_density}
\eeq
where $j$ enumerates the eigenvalues $\lambda_{i,j}$ equal and closest to zero. 
These eigenvalues $\lambda_{i,j}$, as well as the corresponding modes
$\psi_{i,j}(x)$, are also characterized by the $i$-th boundary condition.
Correspondingly, the UV-filtered topological density $q_{i,N}(x)$ depends on 
the boundary condition, too.

We have applied a cluster analysis with a variable lower cut-off $q_{\rm cut}>0$
to these density functions. In a first step the algorithm identifies the interior 
of all clusters (``topological cluster matter'') as the region where 
$|q(x)| > q_{\rm cut}$. The crucial next step is to enquire the connectedness 
between the lattice points in order to form individual clusters out of this 
``cluster matter''. Neighbouring points with $|q(x)|$ above threshold and 
sharing the sign of topological density are said to form the same cluster. 
The cut-off $q_{\rm cut}$ has been chosen such as to resolve the given 
continuous distribution into a maximal number of internally connected while 
mutually separated clusters. The cut-off value has been independently adapted 
for each configuration. The purpose of the cluster analysis was to discover 
extended objects that we are going to consider as dyon candidates.

For the lower temperature ($\beta=8.20$) we have found the following average 
numbers of clusters per configuration comprising all 50 configurations
and corresponding to the boundary conditions $i=1,2,3$:
$$ N_1 = 18.4 (0.4), \quad N_2 = 18.6 (0.5), \quad N_3 = 16.6 (0.3)\,.$$
These numbers (and also the following ones) are ordered in a way which 
corresponds to spatially averaged Polyakov loop values $L$ globally rotated 
into the real $Z(3)$ sector $-\pi/3 < \arg(L) < \pi/3$.  
$N_1$ and $N_2$ are coinciding within errors (the latter given in parentheses),
while $N_3$ is slightly lower. 
The average size of all clusters amounts to 146 lattice points. 
That in the confining phase the abundance of all three types of clusters is 
roughly equal can be interpreted in terms of dyons with maximally nontrivial 
holonomy (see Appendix). Assuming the lattice scale fixed by 
the transition temperature for $SU(3)$ gluodynamics, 
$~T_c = 300~\mathrm{MeV}$~ \cite{Gattringer:2002mr} we obtain the physical
dyon cluster density. It is equal to $6~\mathrm{fm}^{-4}$ and corresponds 
to the presented above $50 \cdot (N_1 + N_2 + N_3)$ clusters. 

Since we observe a slight asymmetry $N_3 < N_1=N_2$ pointing to an 
admixture of configurations related to deconfinement we have applied a 
cut for the modulus $|L| < 0.3$ 
(fixed after improved cooling according to the description given in the
following section). We find from the 23 configurations
surviving this cut the average numbers per configuration 
$$ N_1 = 17.8(0.6), \quad N_2 = 18.3(0.6), \quad N_3 = 17.2(0.5)\,.$$
These numbers overlap within errors as expected in the confinement case. 

The inverse participation ratios (IPR) calculated from the zero modes  
(see Refs.~\cite{Aubin:2004mp,Polikarpov:2005ey,Ilgenfritz:2007xu}) 
for three types of b.c.'s turn out to be  
$$ IPR_1 = 6.8 (0.8),~IPR_2 = 6.4 (1.0),~IPR_3 = 15.2 (1.7),$$
where the statistical errors are given in parentheses.
Having applied the above mentioned cut for $|L|$ we find
$$ IPR_1 = 5.8(0.9),~IPR_2 = 5.0(0.6),~IPR_3 = 8.5(0.6)\,.$$
With the cut taken into account the asymmetry is much less pronounced but
still there. Having the properties of calorons and their constituents in 
mind (see the last paragraphs of the Appendix) we interprete this finding
as a stronger localization of those dyons getting a larger mass when   
approaching the transition temperature from below or moving into the 
deconfinement phase. Those heavier constituents are expected to be 
statistically suppressed, which seems to be already visible, because 
$N_3$ is a bit lower than $N_1$ and $N_2$.   

Next we have checked whether clusters of different type appeared correlated
among each other. We found at the lower temperature (not applying any cut) 
\begin{eqnarray}
{\rm number~of~isolated~clusters~} &=& 1299 \qquad (49 \%)\,,  \nonumber  \\
{\rm number~of~clusters~in~pairs~} &=& ~782 \qquad (29 \%)\,,  \nonumber  \\
{\rm number~of~clusters~in~triplets~} &=& ~597 \qquad (22 \%) \,, \nonumber 
\end{eqnarray}
where the clusters of different type are counted as connected in pairs or 
triplets if the distance was less than two lattice spacings. 
Interpreted in terms of calorons of non-trivial holonomy this means
that we see full caloron-like clusters consisting of three constituents
on one hand and also completely dissolved caloron constituents on the other
hand.   

For the higher temperature ($\beta=8.25$) we have found the following numbers 
of clusters per configuration (on all 50 configurations) corresponding 
to the b.c.'s $~i=1,2,3$:
$$ N_1 = 20.7(0.6), \quad N_2 = 20.6(0.6), \quad N_3 = 17.1(0.4).$$
The average size of these clusters amounts to 172 lattice points.
The density of clusters is higher than at lower temperature: 
$8~\mathrm{fm}^{-4}$. 
In this ensemble we find already a fully developed asymmetry of the 
distribution of the spatially averaged Polyakov loop $L$. Corresponding 
to that, applying the antiperiodic b.c. to the Dirac operator leads 
to a lower number $N_3$ of clusters of the topological density. 
Under these circumstances given by the ensemble average Polyakov loop 
(i. e. the asymptotic holonomy for a would-be exact caloron) the type of 
dyons corresponding to the antiperiodic boundary condition acquires a higher 
action (see Appendix). Thus, one could have expected them to be suppressed 
in the equilibrium ensemble. 
A pronounced asymmetry between the different b.c.'s is seen 
also in the inverse participation ratios that the zero modes acquire: 
$$ IPR_1 = 7.9(0.8),~IPR_2 = 7.2(0.7),~IPR_3 = 24.6(1.8).$$ 
In other words, the zero modes sitting on ``heavy dyons'' (that we found 
as clusters of the topological density defined by the antiperiodic b.c.) 
are localized about three times as much than the zero modes 
sitting on ``light dyons'' (detected using the other two b.c.'s).

Above the deconfinement temperature ($\beta=8.25$) we studied also
the correlation properties of clusters of different type and found them 
organized as follows: 
\begin{eqnarray}
{\rm number~of~isolated~clusters~}    &=& 1600 \qquad (55 \%)\,,   \nonumber \\
{\rm number~of~clusters~in~pairs~}    &=& ~834 \qquad (28 \%)\,,   \nonumber \\
{\rm number~of~clusters~in~triplets~} &=& ~492 \qquad (17 \%).   \nonumber 
\end{eqnarray}
One can conclude that at the temperature above the phase transition the
amount of fully dissociated dyons has much grown relative to the lower 
temperature at the expense of dyons enclosed in clusters of three 
dyons (i.e. complete calorons). 

\section{Polyakov loop portraits of topological clusters}
\label{sec:polyakov-portrait}

\begin{figure*}[!htb]
\centering
a)\hspace{0.3cm} \includegraphics[width=.37\textwidth]{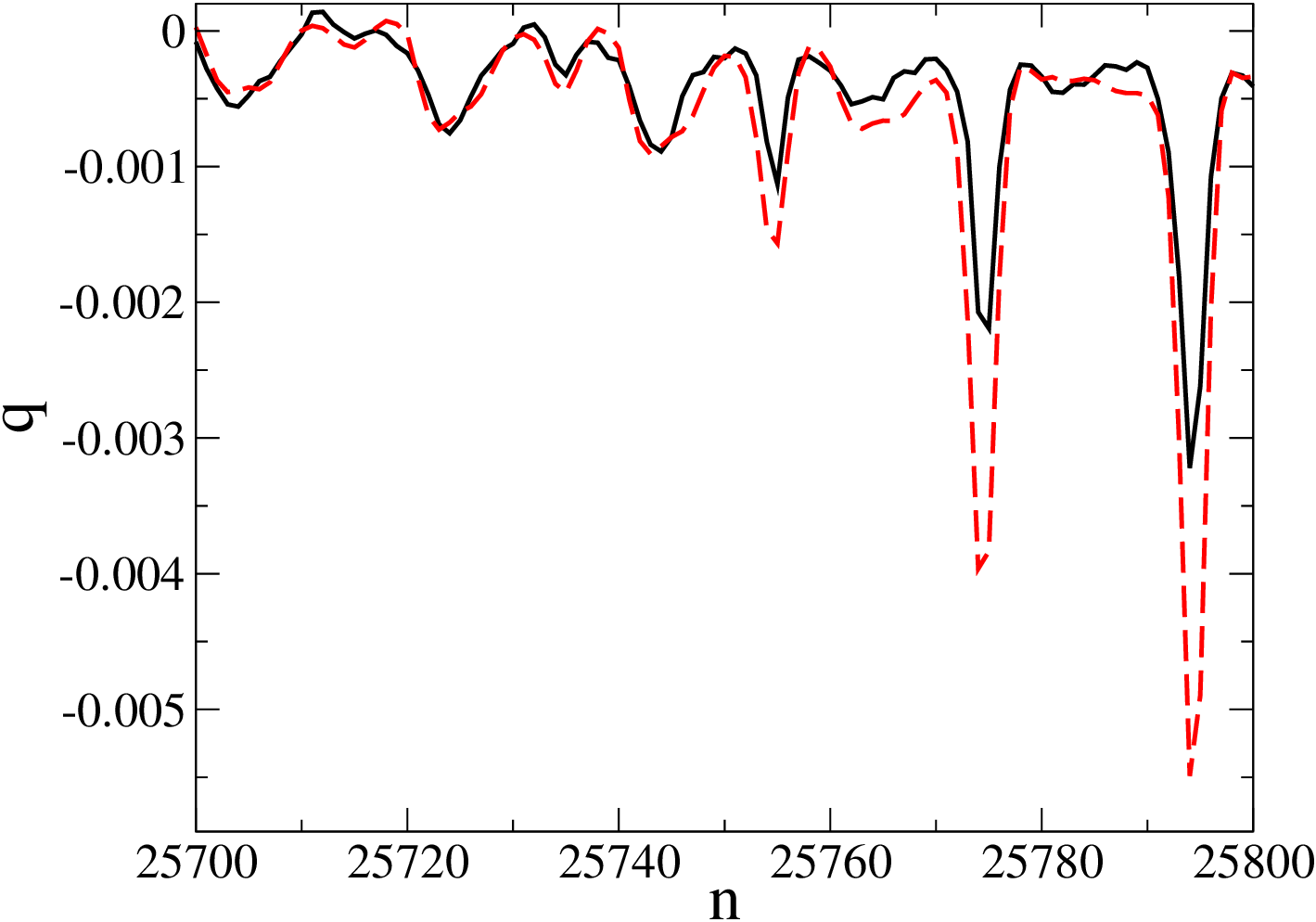}%
\hspace{0.3cm}
b)\hspace{0.3cm} \includegraphics[width=.37\textwidth]{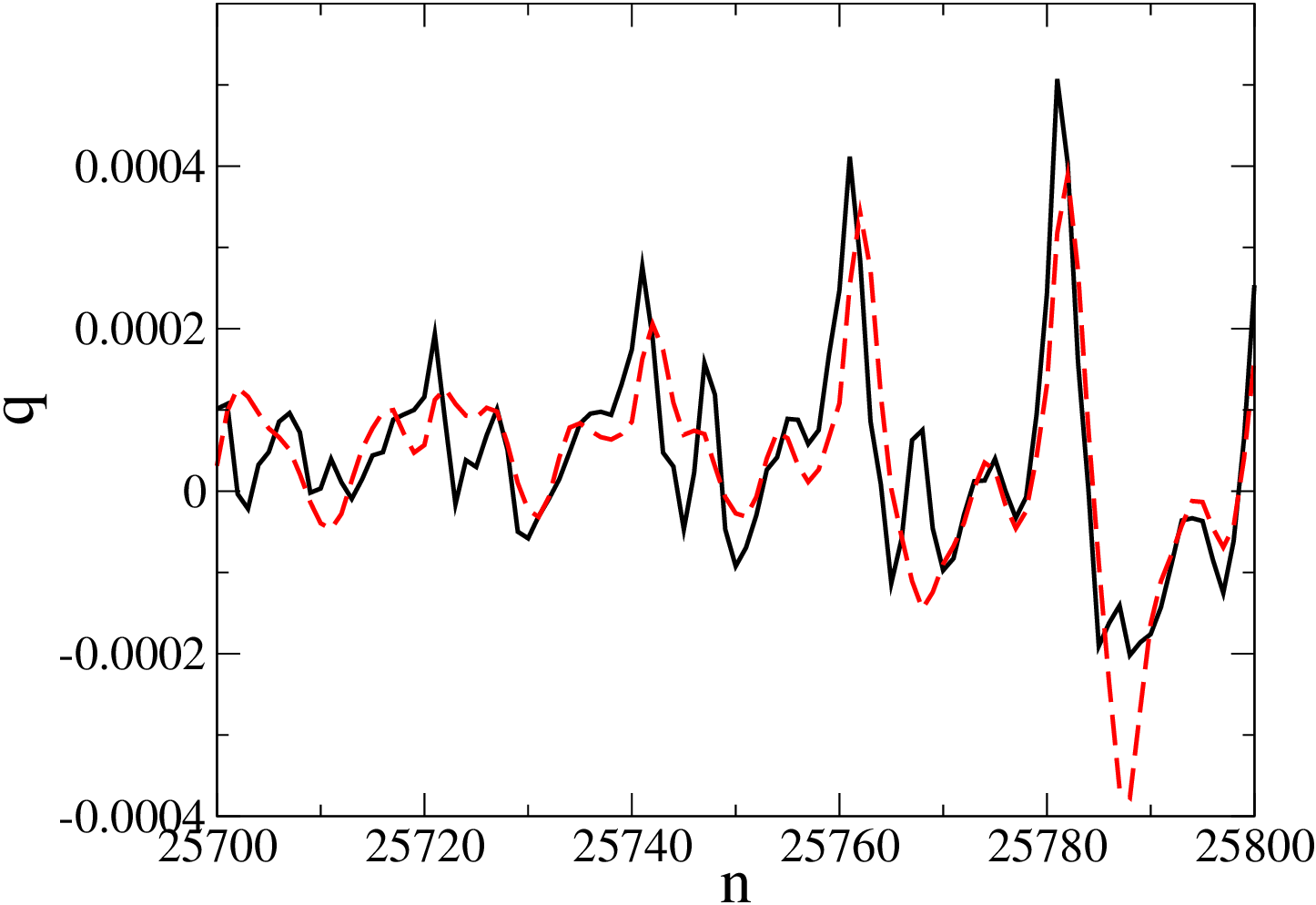}\\
\caption{Comparison (along some string of lattice sites) of the overlap 
topological charge density of a typical configuration for a) $\beta=8.20$ 
and b) $\beta=8.25$ (averaged over b.c.'s, shown as solid black line) with 
the gluonic topological charge density (dashed red line). The latter is 
determined after an optimal number of overimproved cooling steps (see text).}
\vspace{1cm}
\label{fig:comp}
\end{figure*}

In order to provide evidence for the nature of the clusters as dyons
(caloron constituents) the profile of the local Polyakov loop inside 
them should be monitored. 
It is known that some amount of gauge field cooling must be applied to see
emerging patterns of the Polyakov loop. We have used here (as already before)
overimproved cooling~\cite{GarciaPerez:1993ki}. This variant of cooling has 
been found to successfully characterize the gauge field configurations 
according to their total topological charge if it is continued to the final 
(self\-dual or anti-selfdual) plateaux~\cite{Bornyakov:2013iva}.
Here, however, we have restricted ourselves to the first steps of overimproved
cooling. More precisely, we decided for each gauge field configuration 
individually to cool down to the cooling step, when the 
UV-filtered topological charge density of uncooled configurations -- built on
the overlap modes and averaged over the boundary conditions -- is optimally 
fitted by the gluonic topological charge density of the corresponding cooled 
configuration~\cite{Gattringer:2006wq,Bruckmann:2006wf,Bruckmann:2009vb}.%
\footnote{The idea to confront the filtering provided by a given cut-off in 
analyzing modes with the filtering effected by cooling was also the guiding 
idea of Ref.~\cite{Ilgenfritz:2008ia}. Extended to $O(50)$ cooling steps and
matching to the corresponding cut applied to the overlap modes, both techniques 
agreed on the existence of bigger topological lumps with instanton properties.
The search for constituents was not within the scope of that paper.}   
The result of this comparison has lead us to the conclusion that after 
$\approx$ 20 cooling steps a gluonic topological density can be found that 
approaches best the UV-filtered overlap topological density (based on 20 modes 
per configuration). 
The quality of the agreement is demonstrated in Fig. \ref{fig:comp} 
for typical configurations found for $\beta=8.20$ (left panel)
and for $\beta=8.25$ (right panel).  
The characteristic peak structure found along lines of 
subsequent lattice sites (periodically penetrating the lattice several times 
with a minimal off-set) is reproduced by both (fermionic and gluonic) methods.
In the average the agreement turned out to be a bit better for the confinement 
case ($\beta=8.20$) than for deconfinement one ($\beta=8.25$).
Closely related to this, for 90\% (85\%) of the configurations 
obtained at $\beta=8.20$ ($\beta=8.25$) we observed the total gluonic 
topological charge measured after cooling to be equal to the topological
charge assigned by the number of zero modes of the overlap Dirac operator 
without any cooling, i. e. in accordance with the index theorem. 
We should recall that -- for our choice 
of the gauge action and sufficiently high $\beta$-values -- the index was 
found independent of the chosen fermionic boundary condition.

\begin{figure*}[!htb]
\centering
a)\hspace{0.3cm} \includegraphics[width=.37\textwidth]{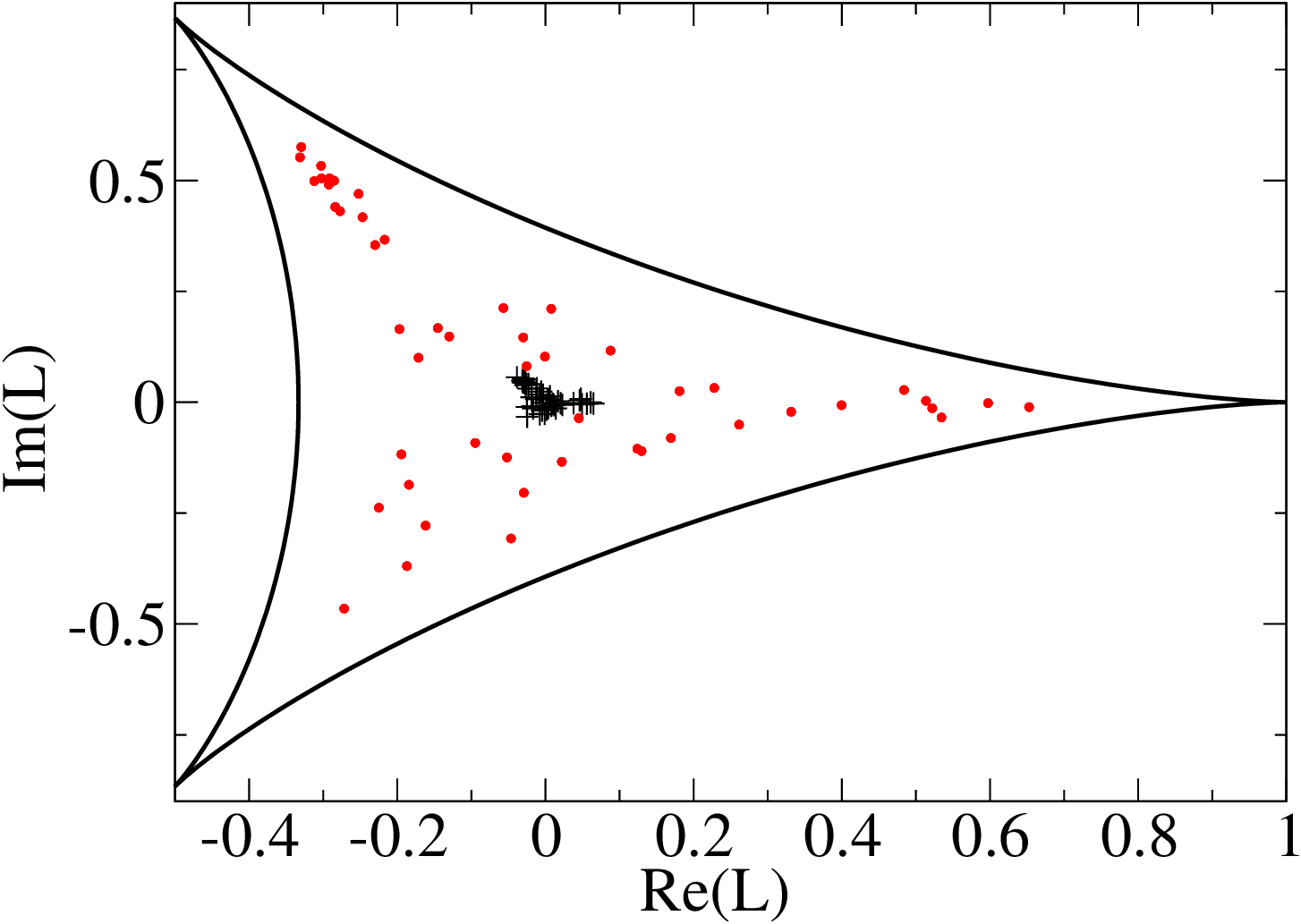}%
\hspace{0.3cm}
b)\hspace{0.3cm} \includegraphics[width=.37\textwidth]{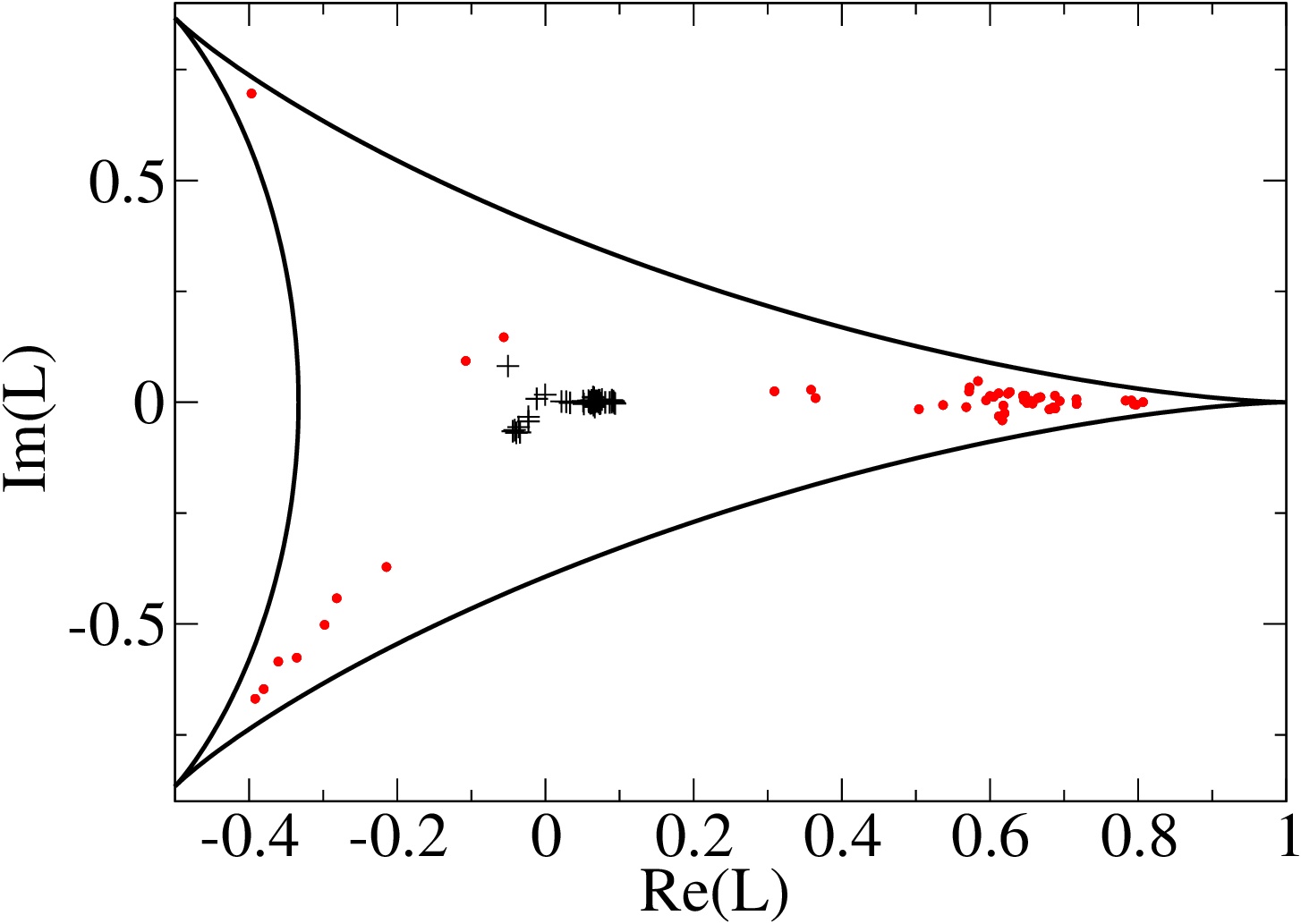}\\
\caption{Scatter plots of the spatially averaged Polyakov loop $L$ before 
(black crosses) and after cooling (red dots)  for a) $\beta=8.20$ 
and b) $\beta=8.25$, respectively.}
\label{fig:plundercooling}
\vspace{1cm}
\end{figure*}

With the cooling steps the local pattern of the Polyakov loop (once it 
has become discernible from the noise) is smoothed~\cite{Bornyakov:2013iva}.
A related effect is reported in Fig. \ref{fig:plundercooling}, 
where the scatter plots of the {\it spatially averaged} Polyakov loop $L$
before and after cooling are shown for a) $\beta=8.20$ and b) $\beta=8.25$, 
respectively. 
For $\beta=8.20$ (confinement) the initial distribution around the origin 
expands more or less isotropically, while in the corresponding scatter plot 
for $\beta=8.25$ (deconfinement) the spatially averaged Polyakov loop $L$ 
is driven by cooling toward the three ``corners'' representing 
trivial holonomy.  

From the study of classical calorons and their dyon 
structure~\cite{Ilgenfritz:2005um} we know that the ``center of the dyon'',
is characterized by a minimal distance between two almost-degenerate 
eigenphases of the local holonomy.

\begin{figure*}[!htb]
\centering
a)\hspace{0.1cm} \includegraphics[width=.27\textwidth]{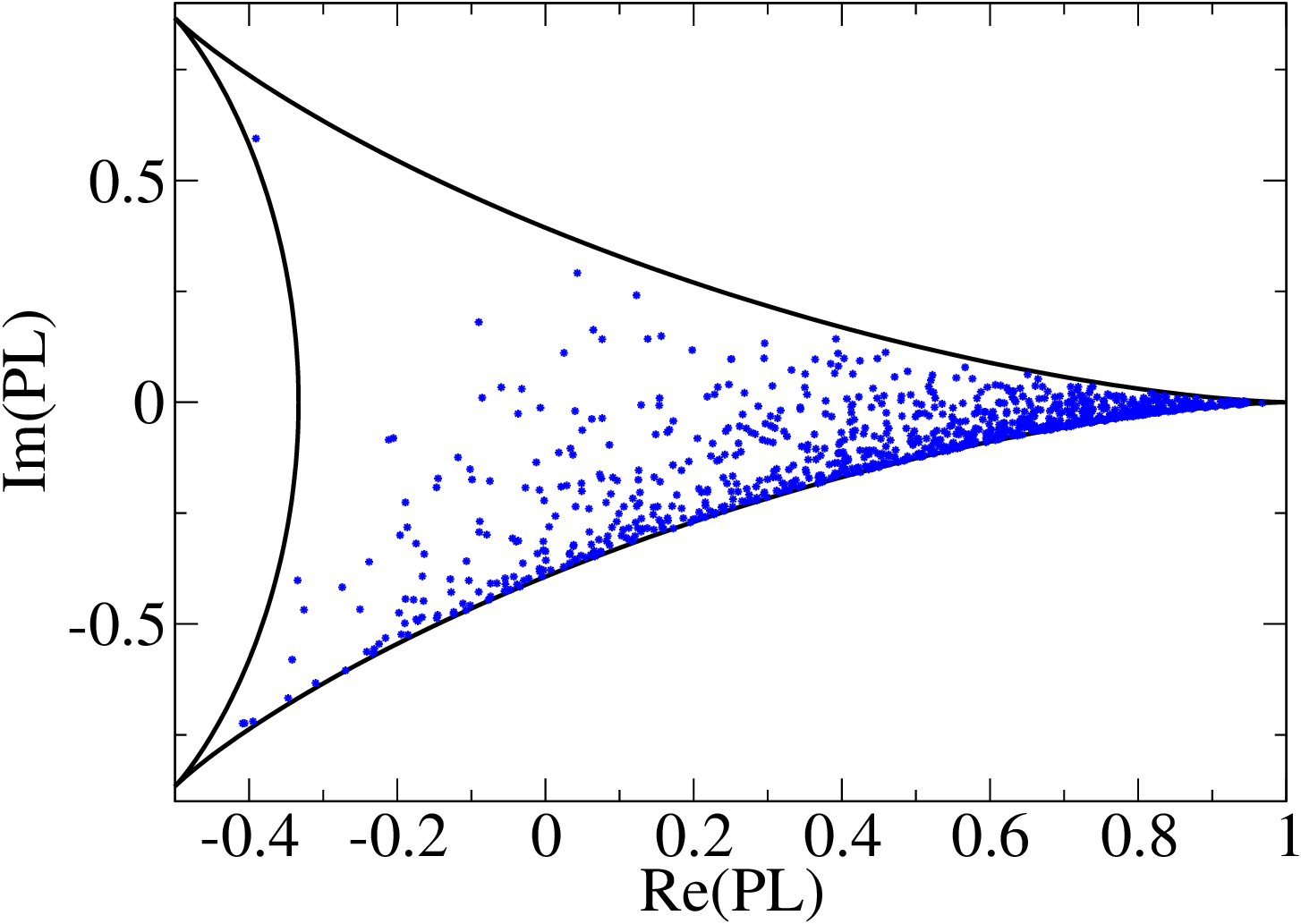}%
\hspace{0.1cm}
b)\hspace{0.1cm} \includegraphics[width=.27\textwidth]{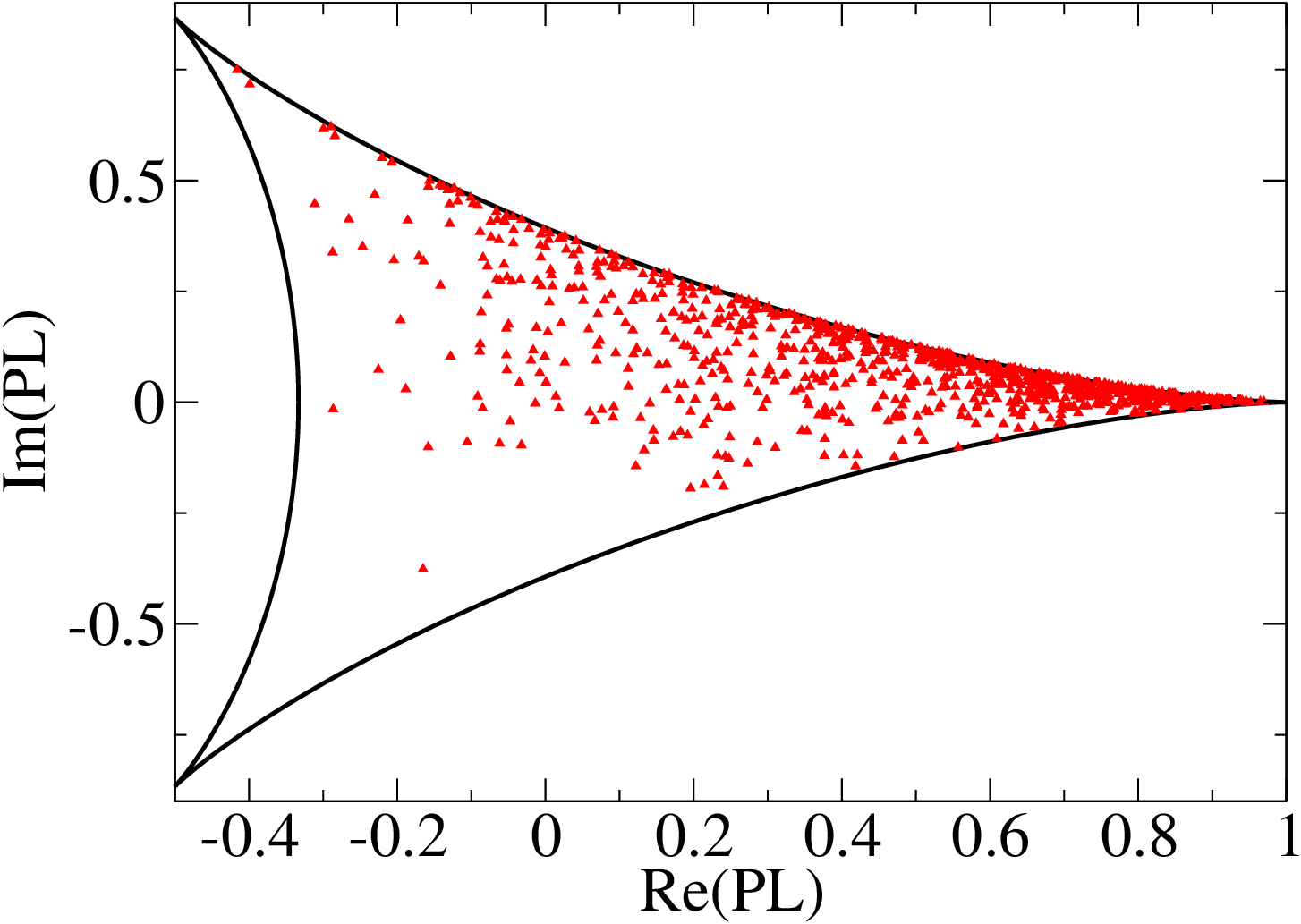}%
\hspace{0.1cm}
c)\hspace{0.1cm}\includegraphics[width=.27\textwidth]{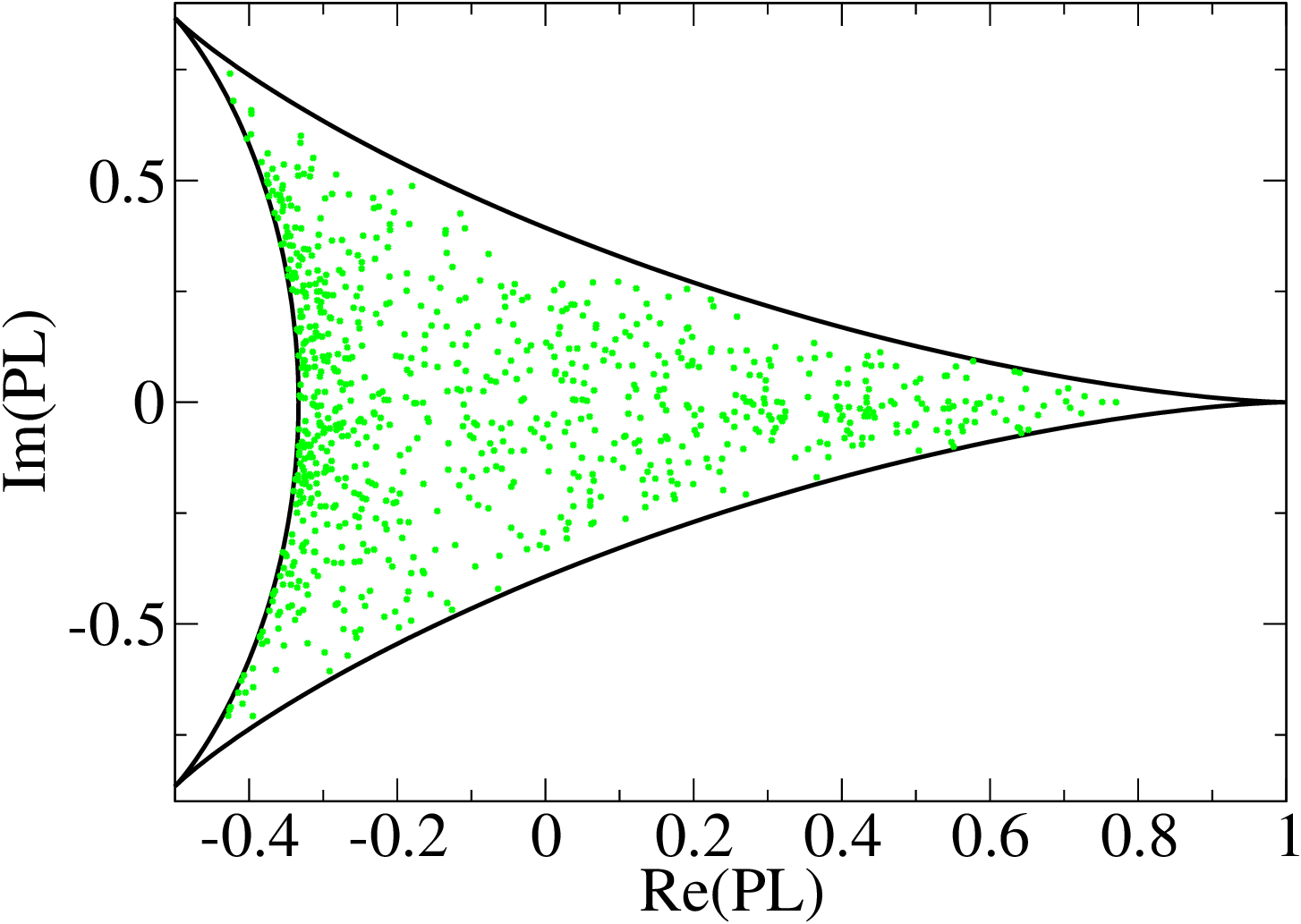}\\
\caption{
For three cases of the fermion temporal b.c., the zero and 
near-zero modes of $SU(3)$ overlap Dirac operator define three different 
profiles of the topological charge density on a configuration belonging 
to the ensemble at $\beta=8.20$ ($T \simeq T_c$). 
Each cluster of these profiles is represented in the plots by the Polyakov 
loop $PL$ measured in its cluster center. The ordering corresponds to
spatially averaged Polyakov loop rotated into the real $Z(3)$ sector 
(after improved cooling). 
a) 918 clusters of first type (blue), 
b) 929 clusters of second type (red),
c) 831 clusters of third type (green).  }
\vspace{1cm}
\label{fig:plfor3typesofdyons_820new}
\end{figure*}

\begin{figure*}[!htb]
\centering
a)\hspace{0.1cm} \includegraphics[width=.27\textwidth]{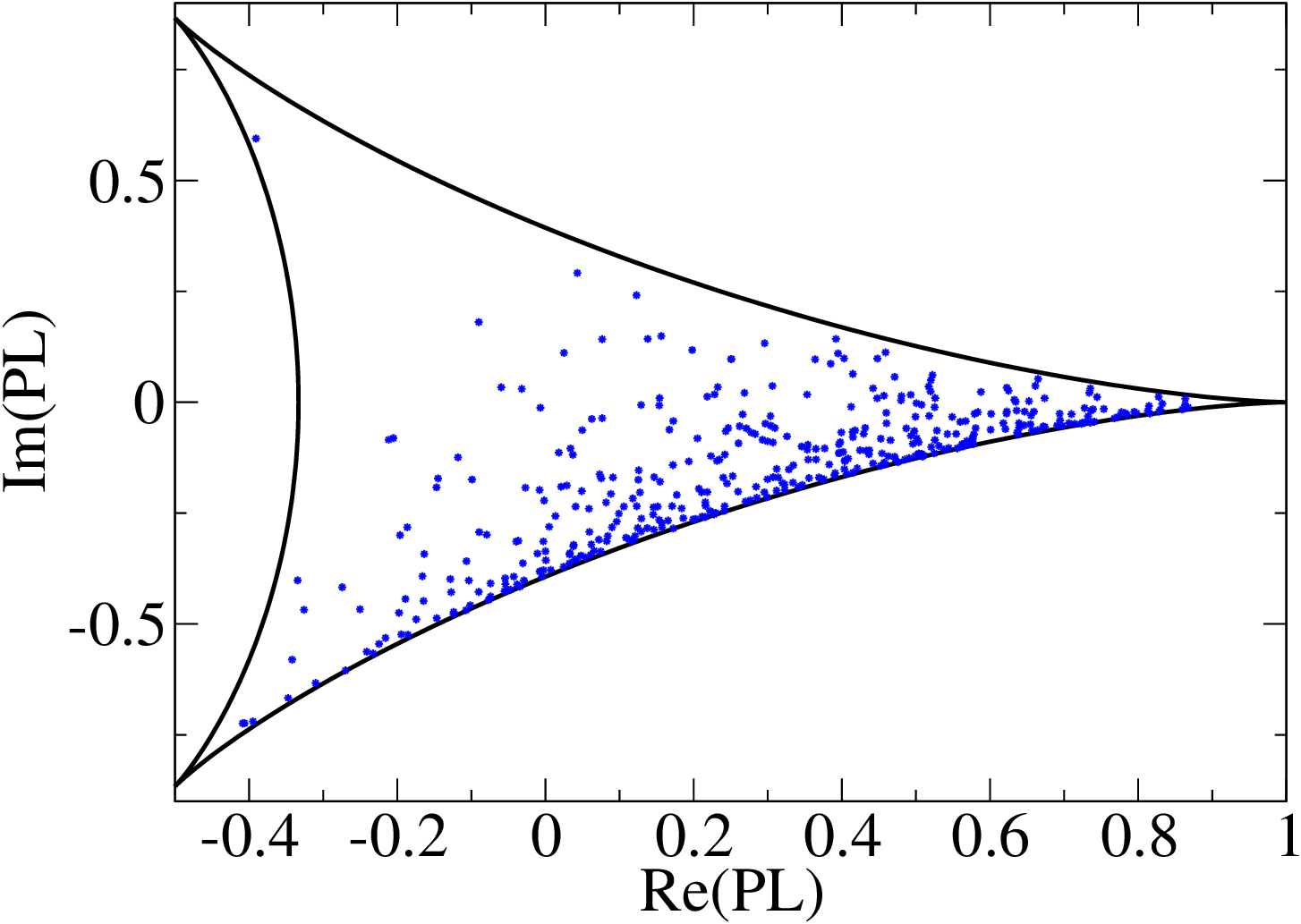}%
\hspace{0.1cm}
b)\hspace{0.1cm} \includegraphics[width=.27\textwidth]{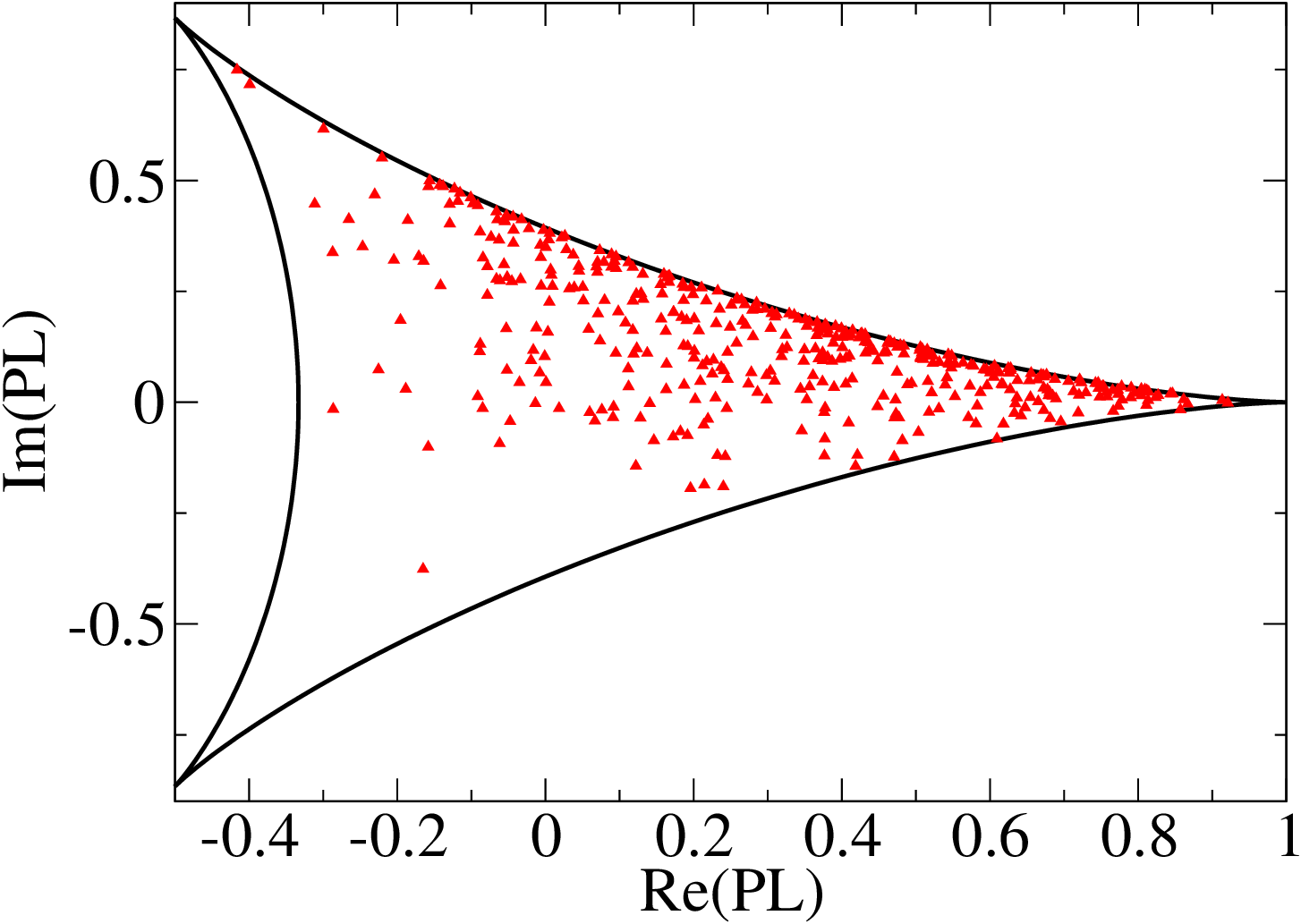}%
\hspace{0.1cm}
c)\hspace{0.1cm}\includegraphics[width=.27\textwidth]{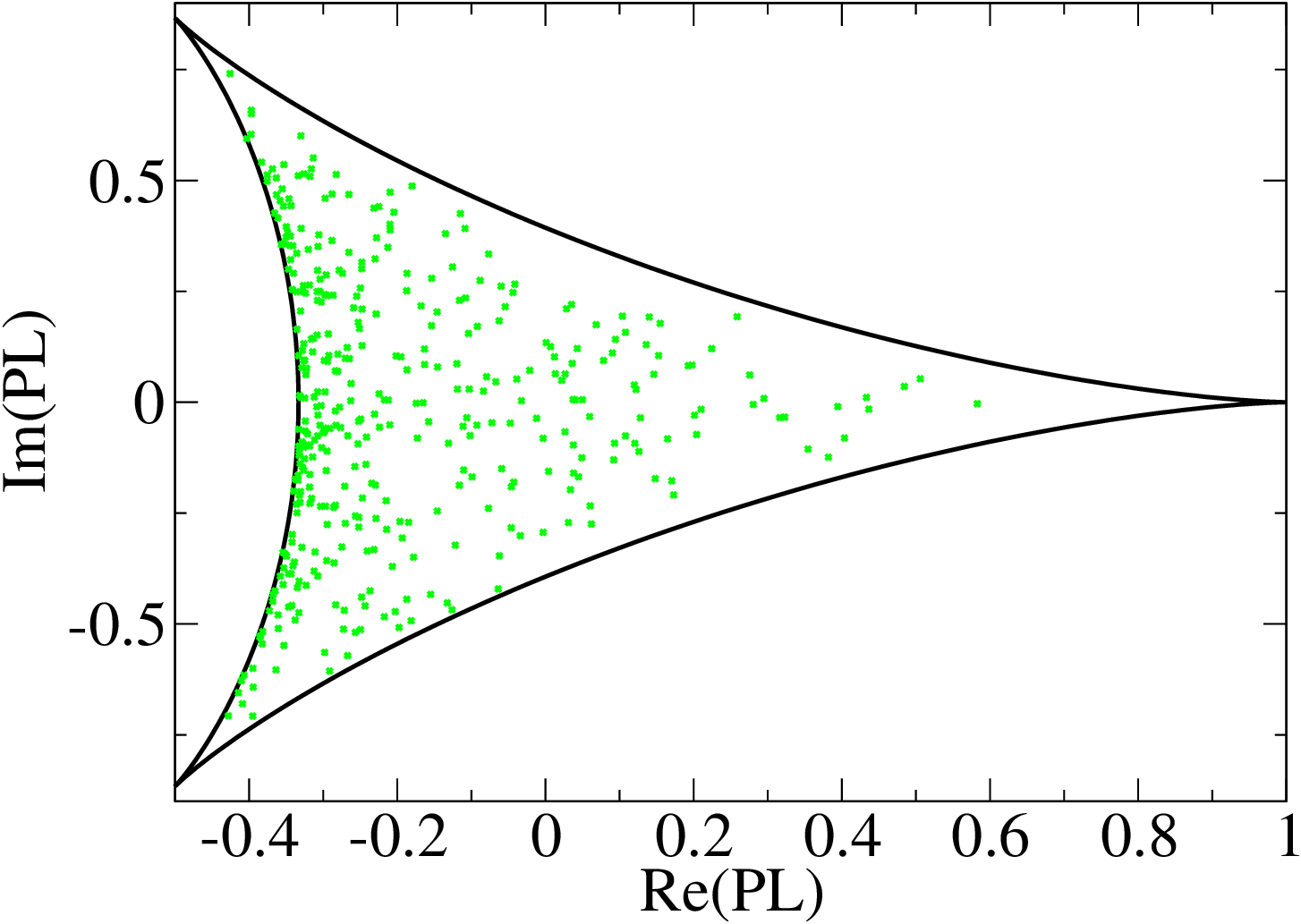}\\
\caption{
Same as in \Fig{fig:plfor3typesofdyons_820new} but with 
a cut $|L|<0.3$ (after improved cooling) leading to 23 (out of 50) 
configurations.
a) 409 clusters of first type (blue),
b) 421 clusters of second type (red),
c) 395 clusters of third type (green).
}
\vspace{1cm}
\label{fig:plfor3typesofdyons_820cut}
\end{figure*}

Having carried out the optimal number of cooling steps (see above) 
the local Polyakov loop - in the following denoted `$PL$' - measured at 
the {\it center of each cluster} (corresponding to b.c.'s 
of type $i=1, 2$ and $3$ and to spatially averaged Polyakov loop values 
$L$ rotated into the real $Z(3)$ sector $-\pi/3 < \arg(L) < \pi/3$)
for $\beta=8.20$ (confinement) has been put into the scatter plots shown 
in Fig. \ref{fig:plfor3typesofdyons_820new}.
The dyonic nature of the clusters would be underlined by the correlation 
between the temporal b.c. applied to the $SU(3)$ overlap 
Dirac operator (selecting one sort of dyons showing up in the cluster 
analysis) and the values of the local Polyakov loop in the center of 
the corresponding topological clusters. What is actually seen in 
Figs. \ref{fig:plfor3typesofdyons_820new}a, b and c is a statistical 
preference of coincidences of two holonomy eigenvalues. 
The scatter plots are relatively concentrated along the appropriate sides 
of the Polyakov triangle (where two eigenphases collide). This happens on 
the side pointing towards the respective angle $\phi$. 
Having in mind that some of the configurations can be related to
deconfinement we have applied the cut $|L|<0.3$ also for this case.
The corresponding modified result is shown in \Fig{fig:plfor3typesofdyons_820cut}.
We see the clustering at the corners of the Polyakov triangle
has disappeared.

\begin{figure*}[!htb]
\centering
a)\hspace{0.1cm} \includegraphics[width=.27\textwidth]{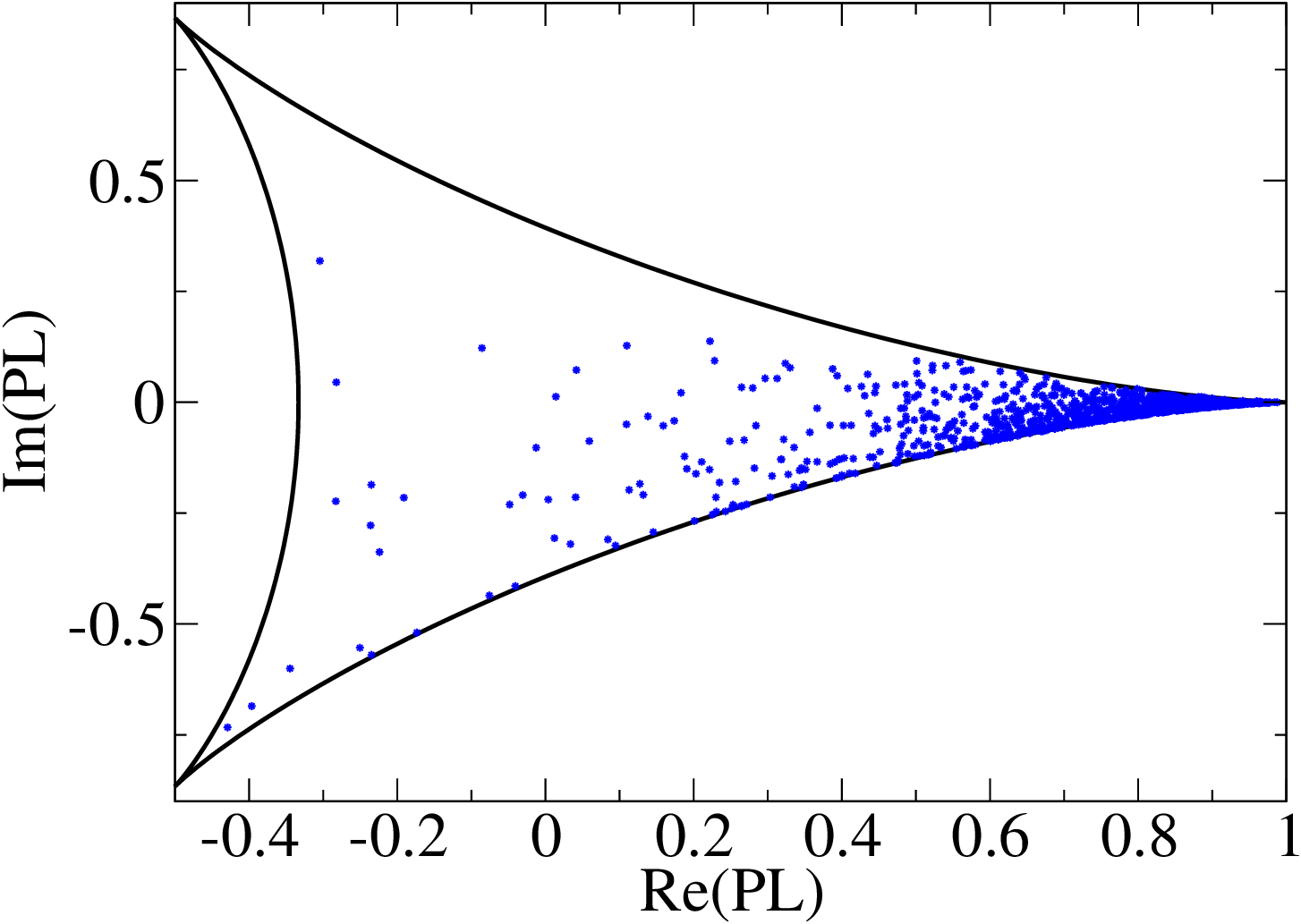}%
\hspace{0.1cm}
b)\hspace{0.1cm} \includegraphics[width=.27\textwidth]{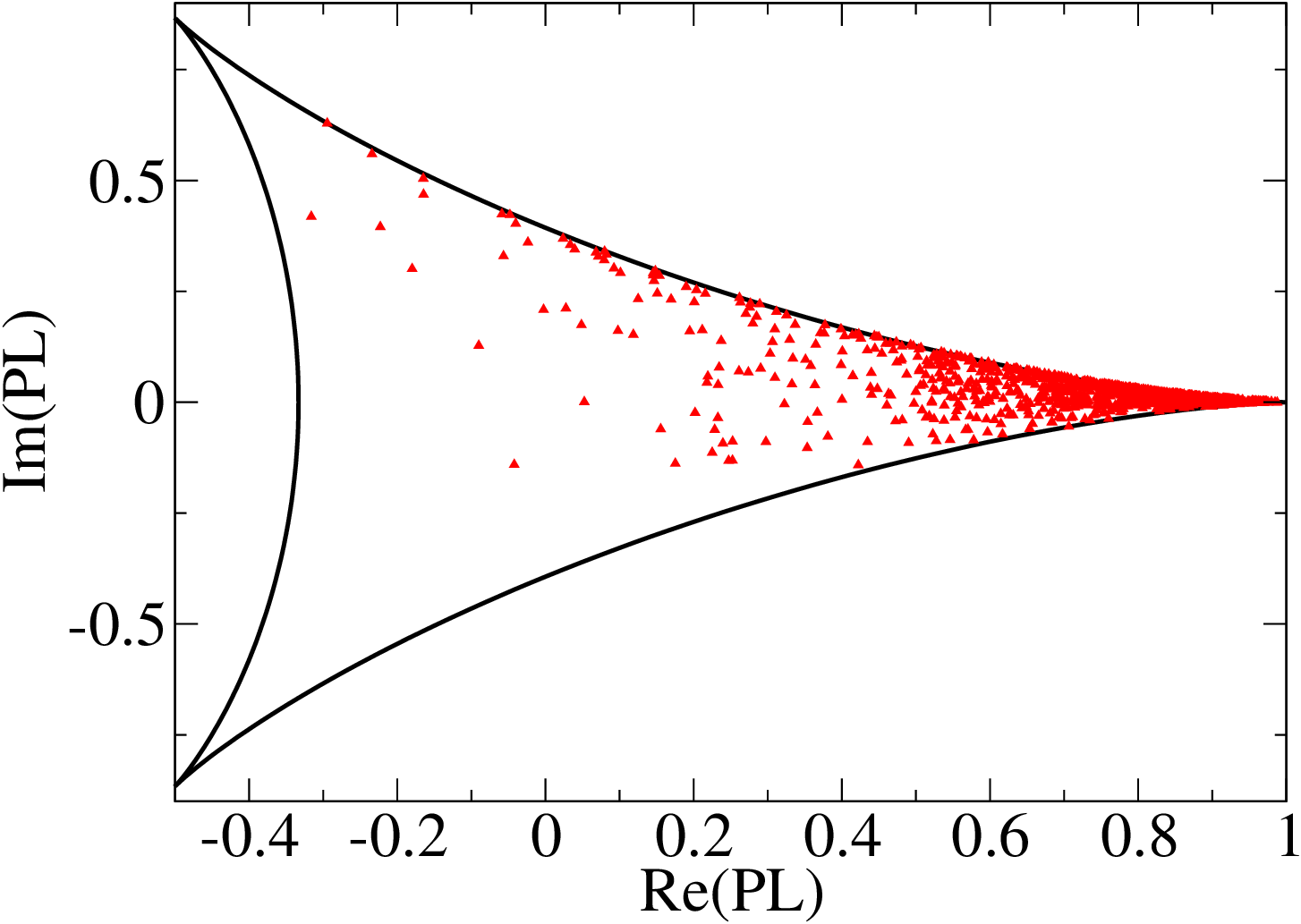}%
\hspace{0.1cm}
c)\hspace{0.1cm}\includegraphics[width=.27\textwidth]{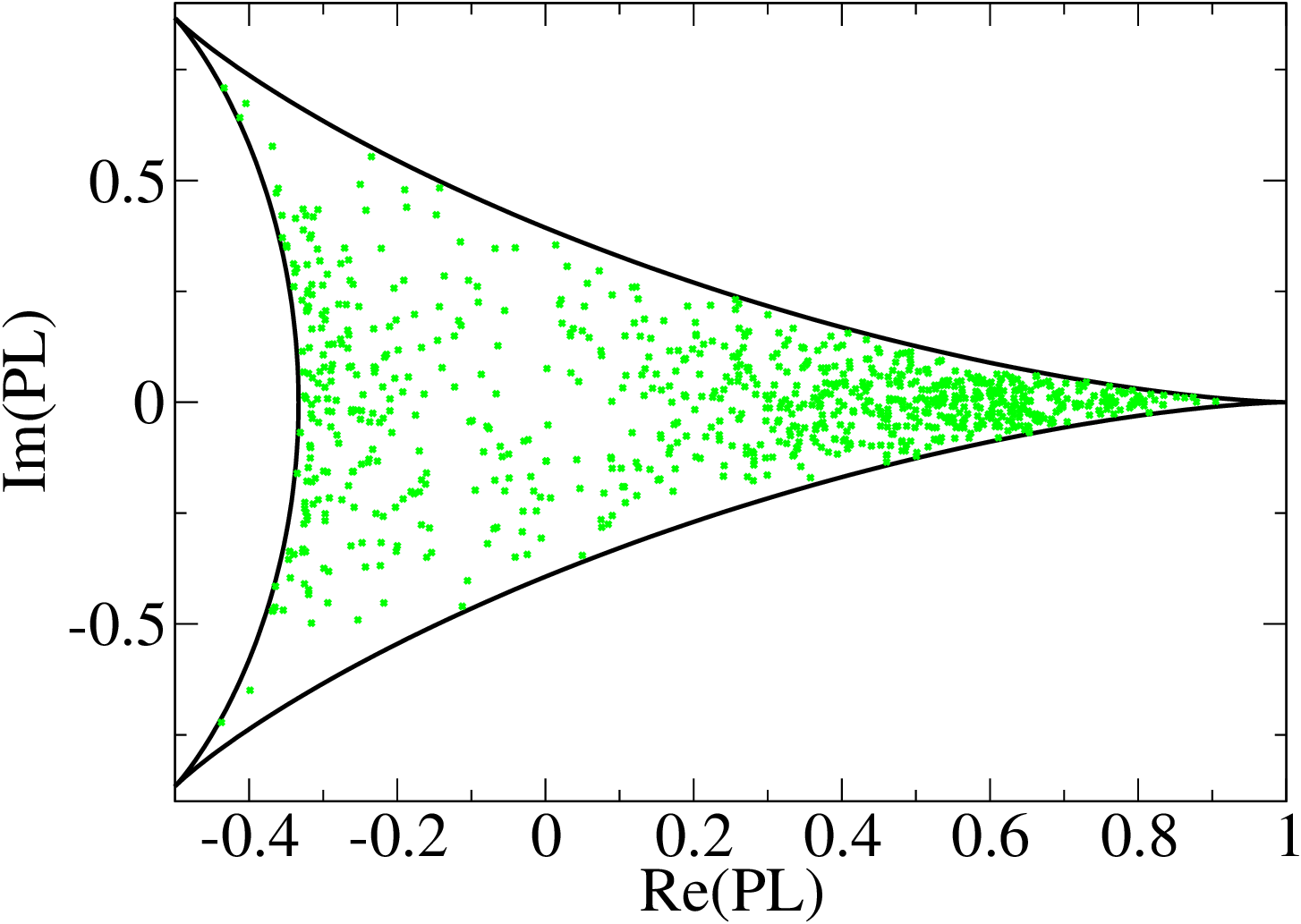}\\
\caption{The same as in Fig. \ref{fig:plfor3typesofdyons_820new} but for
the  ensemble at $\beta=8.25$ ($T \gtrsim T_c$).
a) 1036 clusters of first light type (blue),
b) 1032 clusters of second light type (red),
c) 858 clusters of heavy type (green). }
\vspace{1cm}
\label{fig:plfor3typesofdyons_825new}
\end{figure*}

For $T \gtrsim T_c$, at $\beta = 8.25$, the initially small rotational
asymmetry in the distribution of the spatially averaged Polyakov loop $L$
becomes strongly amplified in the course of cooling in the direction of 
trivial holonomy (see Fig. \ref {fig:plundercooling}b). One sees 
also the small admixture of other (non-real) $Z_3$ sectors (which was 
intentionally hidden by the $Z(3)$ rotation in Fig. \ref{fig:pldistr}a).
The pattern of the local Polyakov loop $PL$ in the centers of the clusters 
is distorted correspondingly, in contrast to the confinement phase, as it 
is shown in \Fig{fig:plfor3typesofdyons_825new}.
Now the scatter plots in \Fig{fig:plfor3typesofdyons_825new}a and b are 
concentrated closer to $PL=+1$, i.e. to the real corner of the Polyakov 
triangle pointing to trivial holonomy. This is what one would expect from
the discussion of KvBLL solutions at the end of the Appendix (see 
\Fig{fig:triangle}). Having a central value $L \simeq +1$ means that those 
dyons observed with b.c.'s $\phi_1$ and $\phi_2$ should have small masses 
$m_1, m_2$ and positions in the Polyakov triangle at the upper and lower
border lines, respectively, close to the corner $PL =+1$.
On the contrary, for antiperiodic b.c. ($\phi_3=\pi$) we would expect to
observe dyons with large mass $m_3$ and positions at the border line
opposite to $PL=+1$. Such objects are really seen in 
\Fig{fig:plfor3typesofdyons_825new}c but only as a minor fraction.
That these heavy objects should be statistically suppressed was expected
from the lowered number $N_3$ in the previous section but not as dramatically
as it is seen here from those points. The majority of clusters in this figure 
are covering points closer to  $PL=+1$. They cannot be explained from 
caloron structures with three constituents each. One may speculate
that they also correspond to (anti)dyons but with some intermediate mass
value. However, we cannot exclude that they are topological fluctuations,
which cannot be interpreted in terms of dyons. They might be also 
``dislocations'' corresponding to some mismatch of the Polyakov loop obtained after
improved cooling and the topological clusters seen from a finite number
of fermion modes.            

\begin{figure*}[!htb]
\centering
a)\hspace{1cm}\includegraphics[width=.27\textwidth]{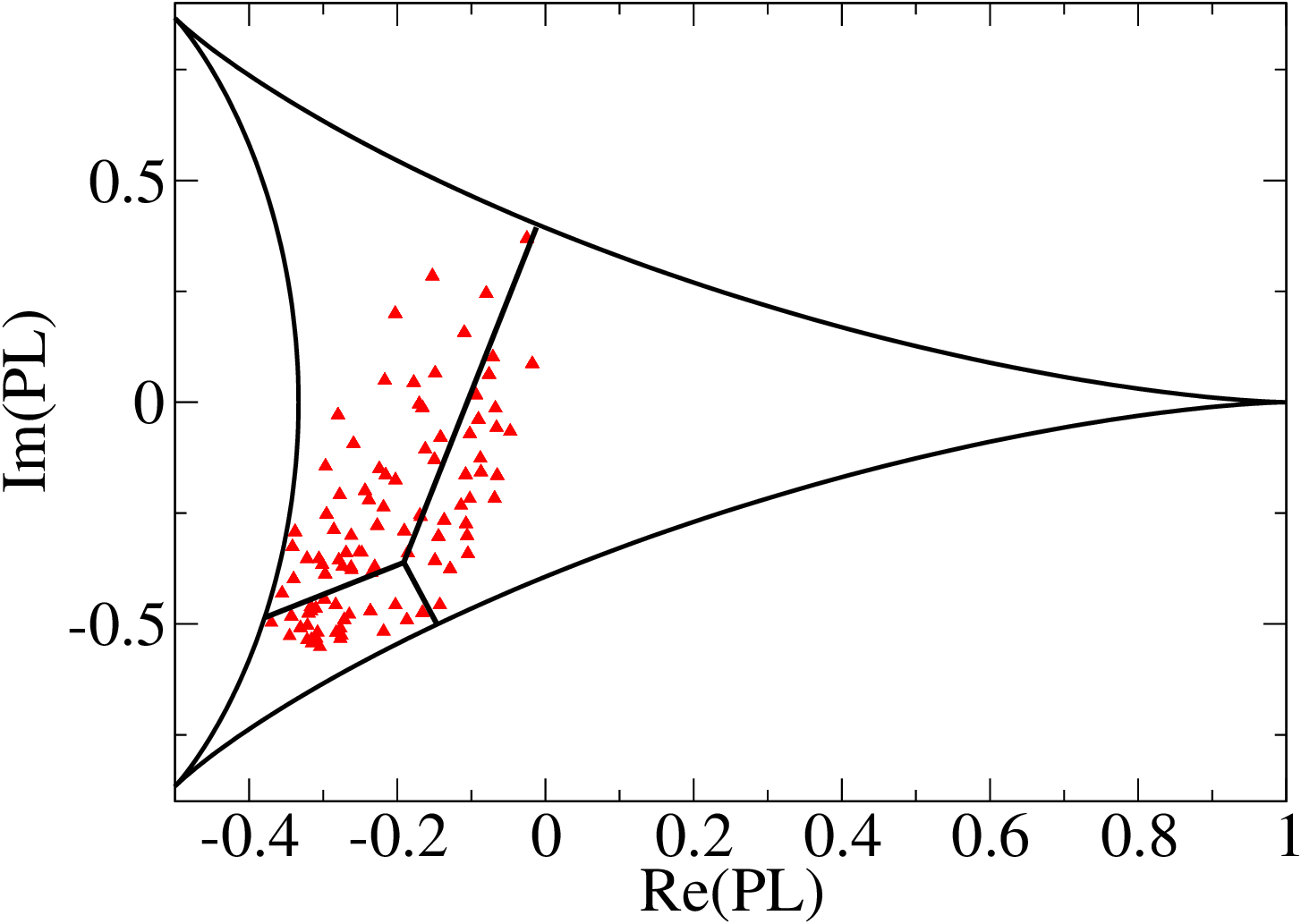}\\
\vspace{1cm}
b)\hspace{0.1 cm}\includegraphics[width=.27\textwidth]{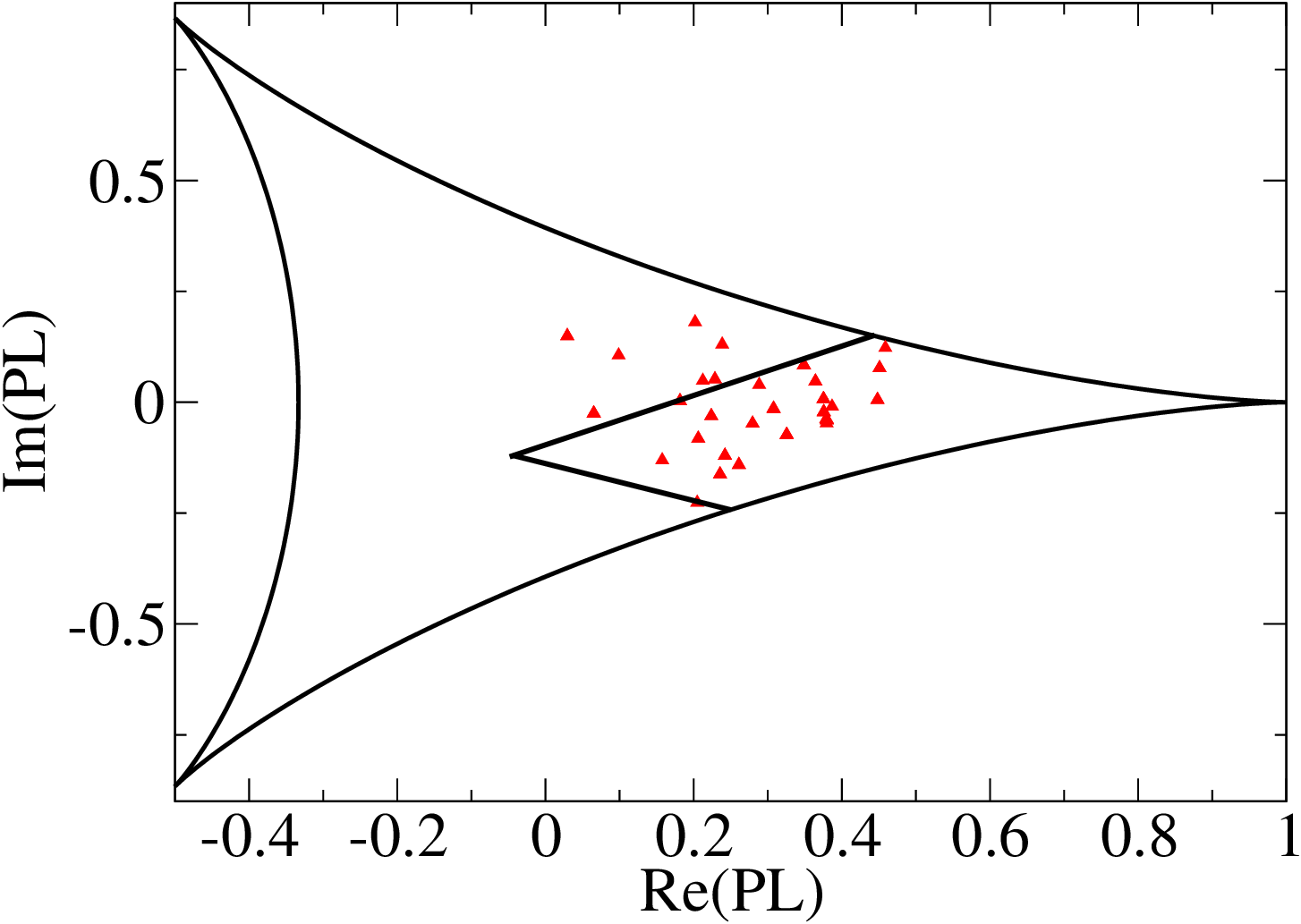}%
\hspace{1cm}
\includegraphics[width=.27\textwidth]{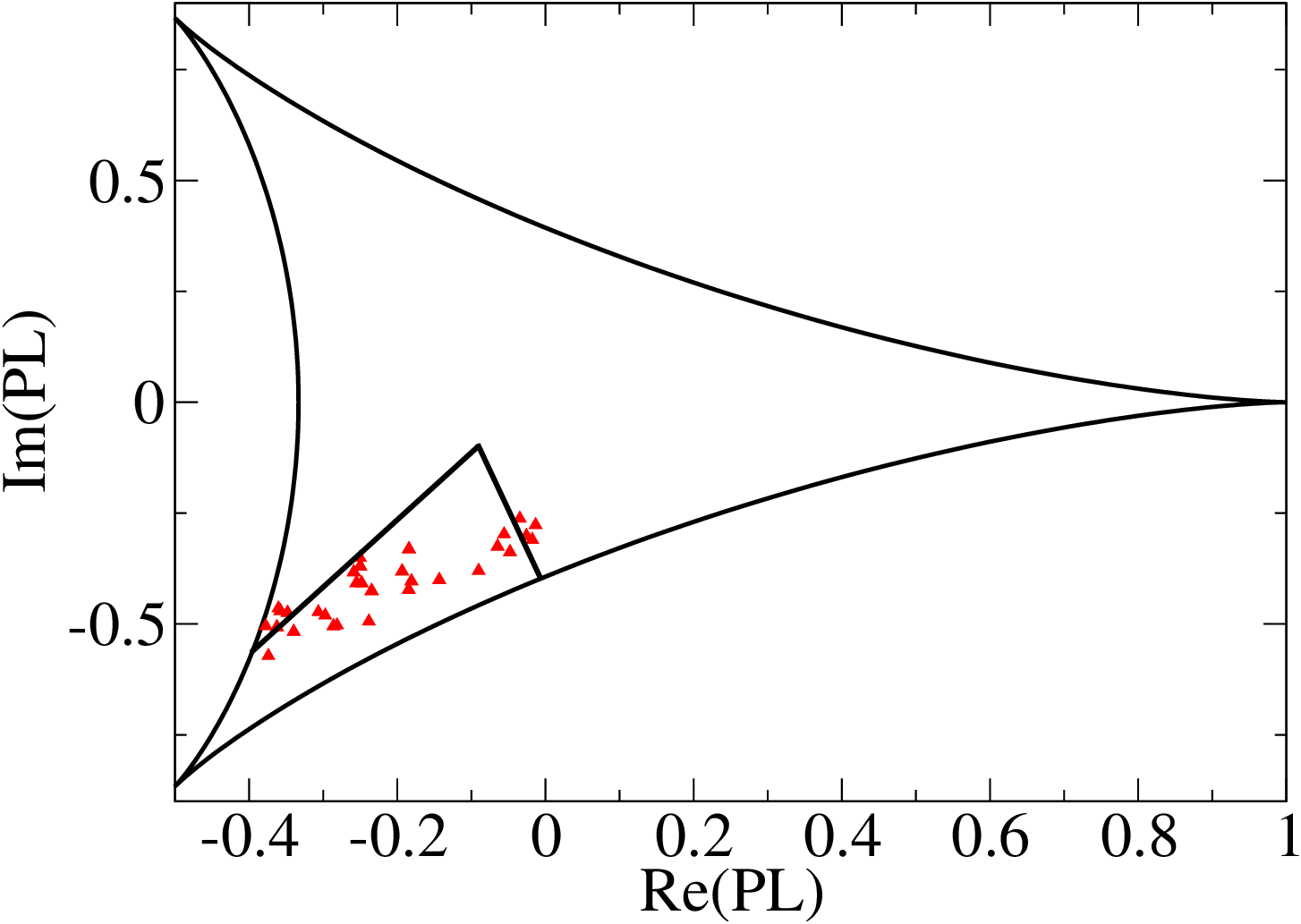}%
\hspace{1cm}
\includegraphics[width=.27\textwidth]{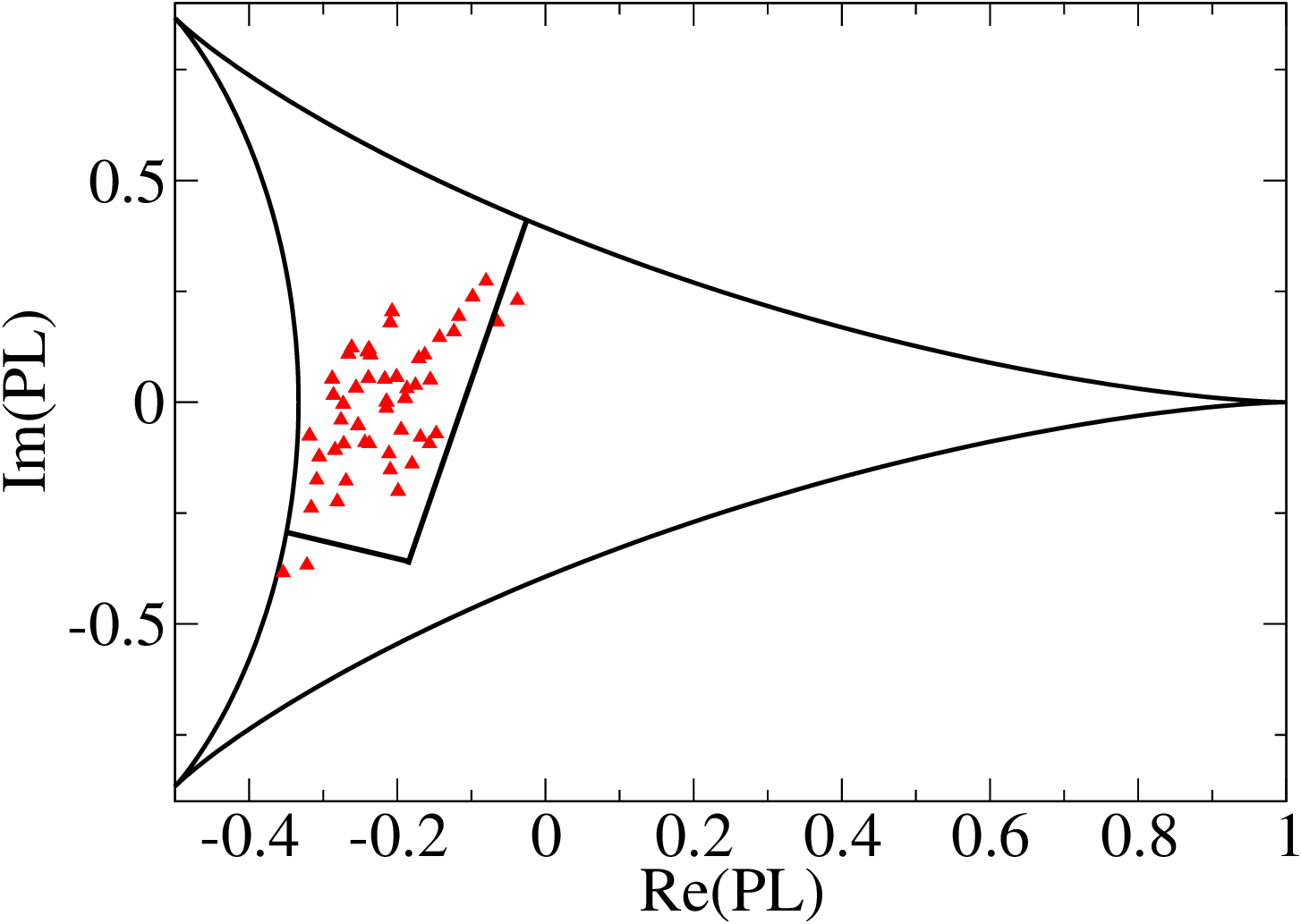}\\
\vspace{1cm}
c)\hspace{0.1 cm}\includegraphics[width=.27\textwidth]{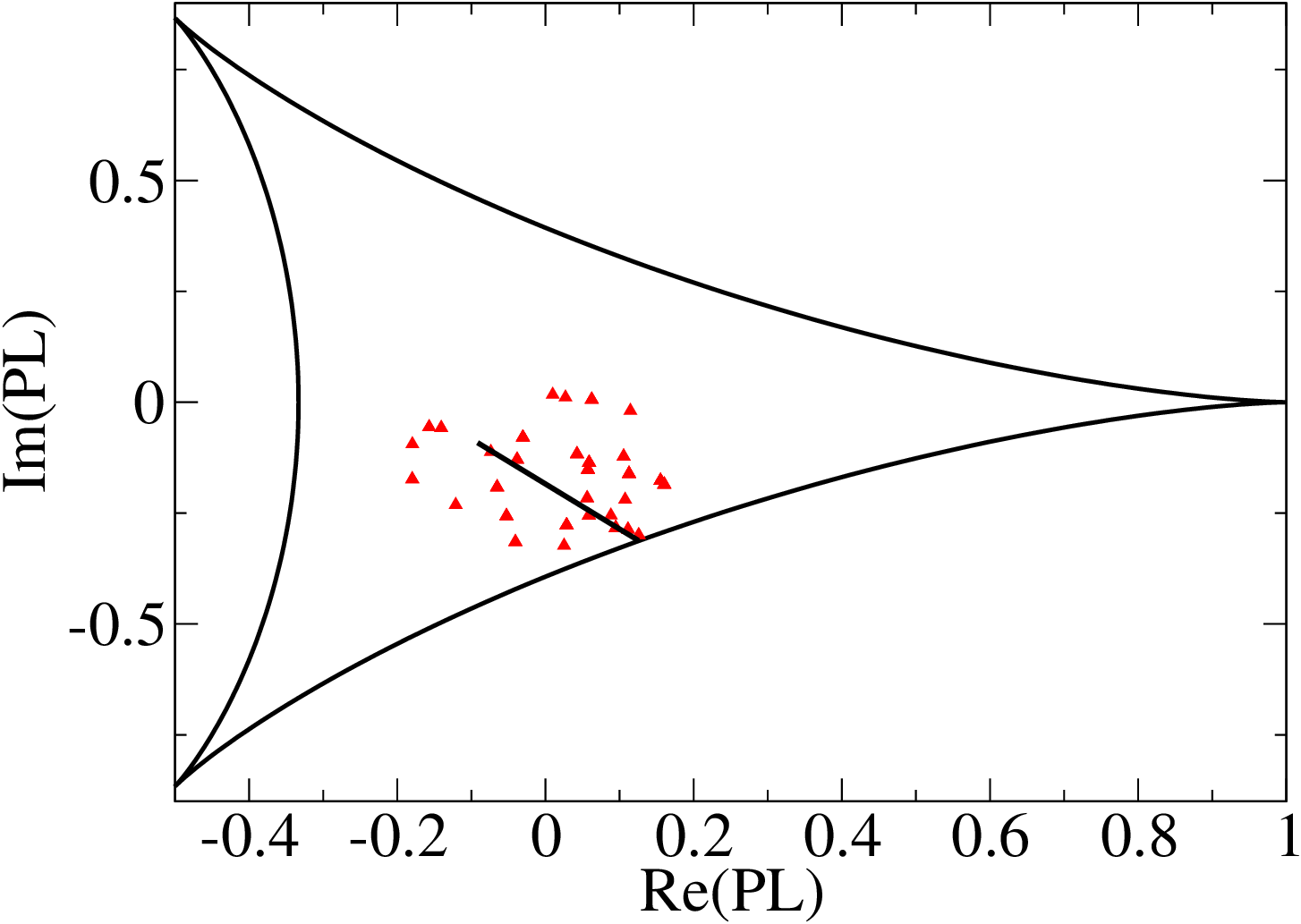}%
\hspace{1cm}
\includegraphics[width=.27\textwidth]{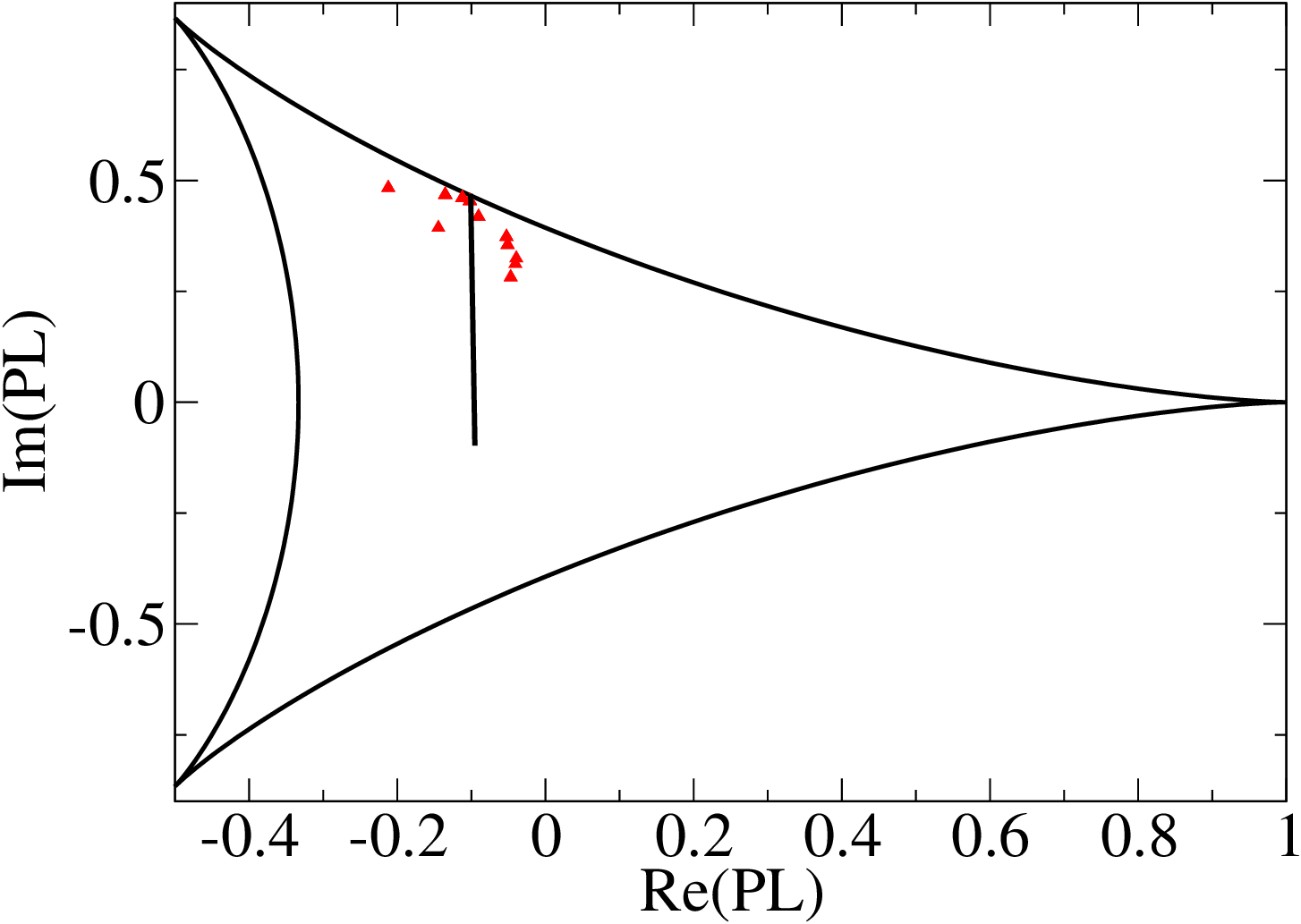}%
\hspace{1cm}
\includegraphics[width=.27\textwidth]{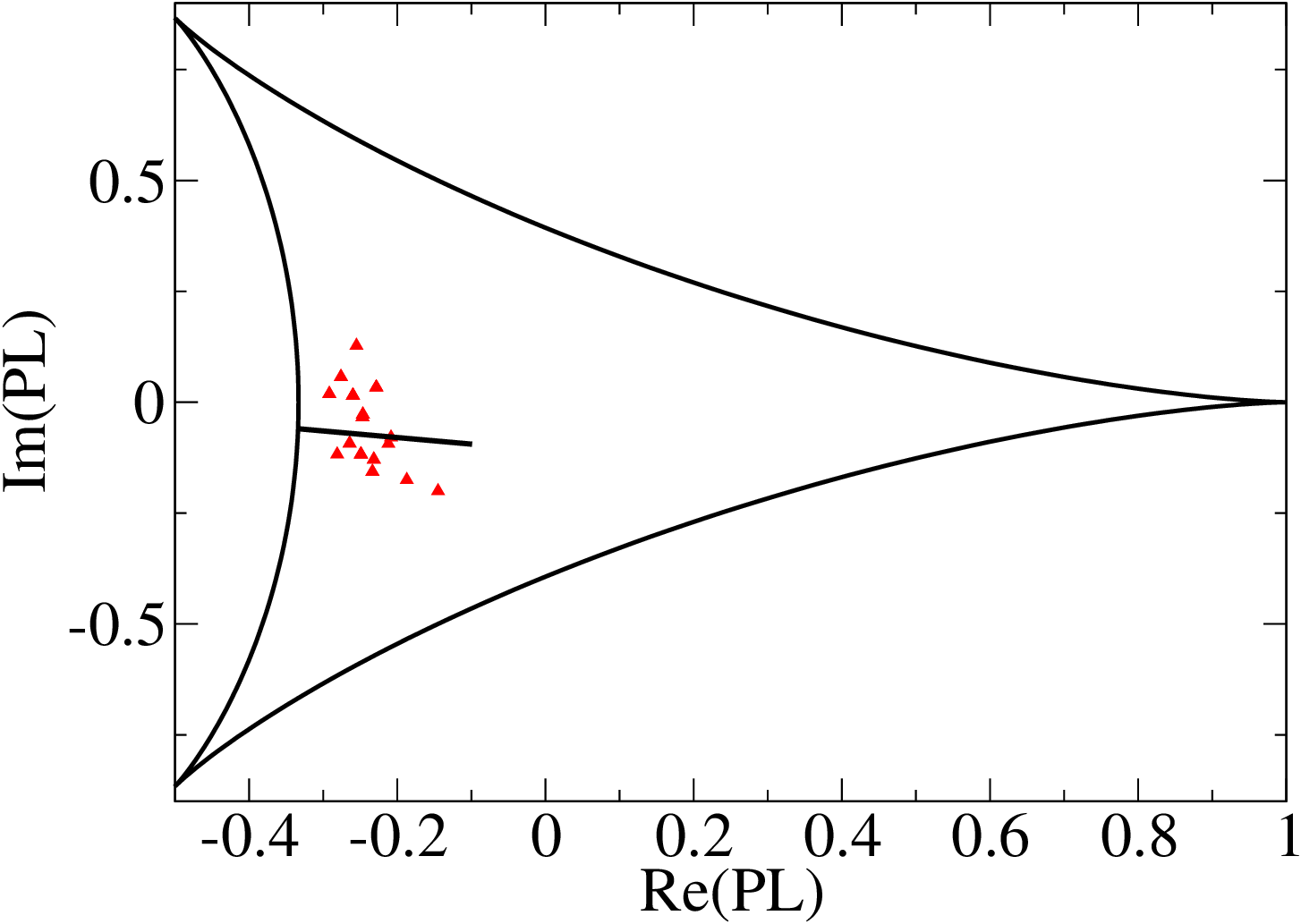}\\
\caption{The profile of the local Polyakov loop $PL$ in examples of 
topological ``clouds'' formed
a) by a triplet of clusters (top),
b) by (three types of) cluster pairs (middle) 
and c) consisting of (three types of) isolated clusters (bottom). 
The pattern obtained for $\beta=8.20$ is characteristic for the 
confining phase. In the figures straight lines are connecting the 
point representing the averaged Polyakov loop $L$ of the cooled 
configuration and the point(s) representing the local Polyakov 
loop $PL$ in the respective dyon center(s).}
\vspace{1cm}
\label{fig:plinclusters}
\end{figure*}

Finally, for both cases $\beta=8.20$ and $\beta=8.25$, respectively 
we have searched for ``clouds'' of topological charge which are either
\begin{itemize}
\item bigger ``clouds'' formed when three dyons of different type are close 
to each other (forming a triplet of clusters $=$ calorons), or
\item bigger ``clouds'' formed when two dyons of different type are close 
to each other (forming three types of cluster pairs), or
\item ``clouds'' consisting of isolated dyons (existing in three types).
\end{itemize}
Scanning the local Polyakov loop point by point, we  are presenting here
examples as scatter plots in Figs. \ref{fig:plinclusters}a, b, c
drawn for the confinement situation ($\beta=8.20$).
The number of points shown corresponds to the actual number of lattice 
points forming the respective example of a topological ``cloud''.
These distributions of points in the complex plane of the Polyakov loop
qualitatively follow the pattern expected for analytical 
calorons~\cite{Gattringer:2003uq,Ilgenfritz:2005um} with their dyonic 
structure. However, this feature is found only in the confining phase, 
whereas in the deconfinement phase the third type of ``clouds'' 
(isolated dyons and possible ``dislocations'') turn out to be dominant. 

\section{Conclusions}
\label{sec:conclusions}

In $SU(3)$ lattice gauge theory, using a small number of modes of the overlap 
Dirac operator with eigenvalues closest to zero, we have investigated clusters 
formed by the (fermionic) topological charge density, appearing in association 
with three different types of boundary conditions that have been applied to the 
fermion field.

Assuming that these clusters correspond to dyons, we have demonstrated how 
their frequency of occurrence and the tendency to combine into triplets 
(calorons) or to form pairs of dyons or to remain isolated depends on the 
temperature. An increasing caloron dissociation is observed as the result 
of the small temperature increase across the phase transition.

It turned out that the angle $\phi$ entering the boundary condition has no
absolute meaning. Instead, the angle only relative to the phase of the average
Polyakov loop gives a unique definition of heavy and light dyons in the deconfined 
phase. Heavy dyons are selected by $\phi$ chosen opposite to the phase of the
average Polyakov loop. Taking the phase of the average Polyakov loop into account 
also in the confined phase does not change the approximate equality of numbers of 
dyons of different type, but it sharpens the relation between the distribution 
of dyons in the complex Polyakov loop plane on one hand and the applied boundary 
condition. In other words, for confinement and deconfinement a unique relation 
between $\phi$ and the positioning of the corresponding dyon clusters in the 
complex plane of the local Polyakov loop can only be achieved if the 
configurations are flipped to the real sector with respect to the averaged 
Polyakov loop.  

Then -- in the confining phase -- the pattern of the Polyakov loop, found inside 
single clusters (``dyons'') of the topological charge densities $q_{i,N}(x)$, 
reflects with high precision the respective fermionic temporal boundary 
condition of type $i$, that has been applied when the low-lying overlap 
spectrum was determined. For dyons within a caloron solution, their monopole 
property {\it strictly requires} that the monopole centers are populating the
sides of the Polyakov triangle. The study of the distribution of the local
Polyakov loop values requires an appropriate number of overimproved cooling 
steps that has been well-defined for the confinement phase.

Just above the transition, cooling strongly influences the averaged Polyakov 
loop as well as the local Polyakov loop inside the topological clusters towards
the trivial corners of the Polyakov triangle (irrespective of the boundary
conditions used to find the clusters).

If -- in the confining phase -- clusters of different type appear correlated (in 
pairs or triples) or isolated, then the Polyakov loop distribution inside the 
topological charge ``clouds'' corresponds to what is expected from the 
classical caloron solutions.

Finally we can conclude that we have confirmed the appearance of topological 
objects like KvBLL calorons and their dyon constituents in the confined phase 
of finite temperature $SU(3)$ Yang-Mills theory at an intermediate resolution 
scale. This is very similar to what we found earlier in the $SU(2)$ 
case~\cite{Ilgenfritz:2006ju,Bornyakov:2007fm,Bornyakov:2008im}.
In the deconfinement phase dominance of light dyon constituents was seen,
whereas the interpretation of the ``heavy'' clusters found with antiperiodic 
boundary conditions did not turn out to be fully conclusive.

In as far the calorons and/or their constituents really may play a major role in 
the path integral representation for the partition function and for providing
e.g. confinement for static quarks can only be seen by model calculations 
like those carried out for the $SU(2)$ 
case~\cite{Gerhold:2006sk,Diakonov:2007nv,Bruckmann:2011yd}.   
Moreover, it remains to be seen, how this picture becomes modified in the 
presence of dynamical fermions, i.e. in full QCD.

\subsection*{Acknowledgments}

B.V.M. appreciates the support of Humboldt University Berlin
where the main part of the work was finalized. We thank F. Bruckmann
for careful reading of the manuscript and some hints to improve it.

\section*{Appendix: $SU(3)$ calorons}
For completeness and better understanding we recall some facts (from 
Ref.~\cite{Bornyakov:2013iva}) and add some more details about calorons 
and their dyon constituents. The $SU(N)$ instantons at finite temperature 
(or calorons) with non-trivial holonomy~\cite{Kraan:1998pm,
Kraan:1998sn,Lee:1998bb} can be considered as composites of $N$
constituent monopoles, seen only when the Polyakov loop at spatial
infinity (holonomy) is non-trivial. In the periodic gauge,
$A_\mu(t\!+\!\cT,\vec x)\!=\!\!A_\mu(t,\vec x)$ it is defined as
\beq
\pl=\lim_{|\vec x|\rightarrow\infty}
P\,\exp(\int_0^\cT A_0(\vec x,t)dt).
\eeq
After a suitable constant gauge transformation, the Polyakov
loop can be characterised by real numbers $\mu_{m=1,...,n}$
($\sum_{m=1}^n\mu_m\!=\!0$) that describe the eigenvalues of the holonomy
\beqa
&&\plo=\exp[2\pi i\,{\rm diag}(\mu_1,\ldots,\mu_n)],\\
&&\mu_1\leq\ldots\leq\mu_n\leq\mu_{n+1}\!\equiv\!1\!+\!\mu_1.\nonumber
\eeqa
In units, where the inverse temperature $\cT=1$,
a simple formula for the $SU(N)$ action
density can be written~\cite{Kraan:1998pm,Kraan:1998sn} :
\beqa
&&\Tr F_{\mu\nu}^{\,2}(x)=\partial_\mu^2\partial_\nu^2\log\psi(x),\\
&&\psi(x)=\half\tr(\cA_n\cdots \cA_1)-\cos(2\pi t),\nonumber\\
&&\cA_m\equiv\frac{1}{r_m}\left(\!\!\!\bea{cc}r_m\!\!&|\vec y_m\!\!-\!
\vec y_{m+1}|\\0\!\!&r_{m+1}\eea\!\!\!\right)\left(\!\!\!
\bea{cc}c_m\!\!&s_m\\s_m\!\!&c_m\eea\!\!\!\right),\nonumber
\eeqa
with $r_m\!=\!|\vec x\!-\!\vec y_m|$ and $\vec y_m$ being the center of
mass radii of $m$ constituent monopoles, which can be assigned a mass
$8\pi^2\nu_m$, where $\nu_m\!\equiv\!\mu_{m+1}\!-\!\mu_m$. Furthermore,
$c_m\!\equiv\! \cosh(2\pi\nu_m r_m)$, $s_m\!\equiv\!\sinh(2\pi\nu_m r_m)$,
$r_{n+1}\!\equiv\! r_1$ and $\vec y_{n+1}\!\equiv\!\vec y_1$.

For $SU(3)$ calorons we correspondingly pa\-ra\-me\-tri\-ze the asymptotic
holonomy as
$\plo =
\mathrm{diag}(\mathrm{e}^{2\pi i\mu_1},\mathrm{e}^{2\pi i\mu_2},
\mathrm{e}^{2\pi i\mu_3})$,
with $\mu_1 \leq \mu_2 \leq \mu_3 \leq \mu_4 = 1+\mu_1$
and $\mu_1+\mu_2+\mu_3 = 0$.  Let $\vec{y}_1$, $\vec{y}_2$ and $\vec{y}_3$ be
three $3D$ position vectors of dyons remote from each other. Then
a caloron consists of three lumps carrying the instanton action split
into fractions $m_1 = \mu_2 - \mu_1$, $m_2 = \mu_3 - \mu_2$ and $m_3 = \mu_4 - \mu_3$,
concentrated near the $\vec{y}_i$.

Provided the constituents are well separated, 
the local holonomies at the constitents' positions 
$~\vec{y}_m,~m=1,2,3~$ are ~\cite{vanBaal:1999bz}
\beqa
\nonumber
\Pol(\vec{y}_1)&=
&\diag(\hphantom{-}e^{-\pi{i}\mu_3},\hphantom{-}e^{-\pi{i}\mu_3},\hphantom{-}
e^{2\pi{i}\mu_3}),\\  
\label{eqn:polinterplay}
\Pol(\vec{y}_2)&=
&\diag(\hphantom{-}e^{2\pi{i}\mu_1},\hphantom{-}e^{-\pi{i}\mu_1},\hphantom{-}
e^{-\pi{i}\mu_1}),\\
\nonumber
\Pol(\vec{y}_3)&=
&\diag(-e^{-\pi{i}\mu_2},\hphantom{-}e^{2\pi{i}\mu_2},-e^{-\pi{i}\mu_2}).
\eeqa

The complex numbers representing the trace of the holonomy, i.e. the Polyakov 
loop $PL = \frac{1}{3}\rm {Tr}\Pol$, occupy some region on the complex plane.
The particular holonomies in Eq. (\ref{eqn:polinterplay}) (with 
twofold degerate eigenvalues) would guarantee that the local Polyakov loops 
at the constituents' positions fall onto the boundary of the complex domain 
(see Figs. \ref{fig:plfor3typesofdyons_820new}, 
\ref{fig:plfor3typesofdyons_820cut}, and \ref{fig:plfor3typesofdyons_825new}). 

The trace of the holonomy $\plo$ is three times the Polyakov loop $PL_{\infty}$ 
close to the spatial average of the Polyakov loop $L$, which is vanishing in the 
confining phase and deviating from zero in the deconfining phase.

\begin{figure*}[t]
\includegraphics[width=.2\textwidth]{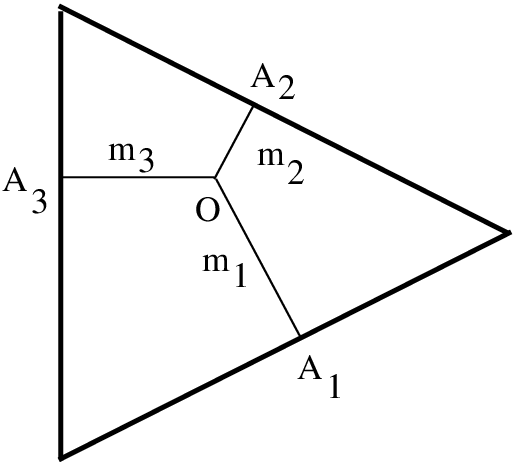}
\caption{ The equilateral triangle with an inner point $O$ and three 
perpendicular projections to the triangle sides. The sum of them 
$OA_1+OA_2+OA_3\equiv m_1 + m_2 + m_3 = 1$ is constant for all inner points.
}
\label{fig:triangle}
\end{figure*}

There is a one-to-one correspondence between the complex Polyakov loop and 
three real numbers (fractions) $m_1, m_2, m_3, m_1 + m_2 + m_3 = 1$ that
are defined by the eigenvalues of $SU(3)$ matrix and hence by its trace. 
Three such numbers can be represented by the inner point of an equilateral 
triangle for which the sum of the lengths of the three perpendiculars to the
triangle sides is constant (equal to one, see Fig. \ref{fig:triangle}). 
The domain in the complex plane occupied by the Polyakov loop can be 
considered as some nonlinear deformation of this equilateral triangle. 
The point $O$ on Fig. \ref{fig:triangle} corresponds to $PL_{\infty}$, while 
the points $A_1, A_2, A_3$ correspond to the values of the local 
Polyakov loop at the constituent positions (\ref{eqn:polinterplay}).
The fermion zero mode $\psi$ in the background of a KvBLL solution 
was found~\cite{GarciaPerez:1999ux,Chernodub:1999wg}
for arbitrary temporal boundary conditions $\psi(t + 1/T, \vec{x}) =
\exp(i  \phi) \psi(t, \vec{x})$. The exciting property of the zero mode 
is that it locates on only one of the dyons. In particular it is located 
on the $m$-th dyon when $\phi/(2\pi) \in \{\mu_m,\mu_{m+1}\}$. 

For the case of maximally nontrivial holonomy expected to be realized 
deeply in the confinement phase we have degenerate masses 
$m_1 = m_2 = m_3 = 1/3$ and $\mu_1 = -1/3, \mu_2 =0, \mu_3 = 1/3$. 
Therefore, when $\phi/(2\pi) \in \{-1/3,0\}$, the zero mode is located 
on the first dyon and for $\phi = \phi_1 = -\pi/3$ its localization
is maximal. For  $\phi = \phi_2 = \pi/3$ and  $\phi = \phi_3 = \pi$ 
it is maximal on the second and third dyon, respectively.

On the contrary, for the limiting case of maximally trivial holonomy 
with $\mu_1=\mu_2=\mu_3=0$ (i.e. $PL_{\infty} = +1$) --
realized for the real $Z(3)$ Polyakov loop sector deeply in the 
deconfinement phase --  we obtain asymmetric constituent masses as
$m_1=m_2=0,m_3=1$. The localization of the only massive constituent is 
then observed with the zero mode related to the antiperiodic boundary
condition ($\phi = \phi_3 = \pi$).         


\bibliographystyle{apsrev}


\end{document}